# Seven Years of Imaging the Global Heliosphere with IBEX

D.J. McComas[1], E.J. Zirnstein[1], M. Bzowski[2], M.A. Dayeh[3], H.O. Funsten[4], S.A. Fuselier[3,5], P.H. Janzen[6], M.A. Kubiak[2], H. Kucharek[7], E. Möbius[7], D.B. Reisenfeld[6], N.A. Schwadron[7], J.R. Szalay[3], J.M. Sokół[2], M. Tokumaru[8]

[1]Department of Astrophysical Sciences, Princeton University, Princeton, NJ 08544, USA (dmccomas@princeton.edu)

[2]Space Research Centre of the Polish Academy of Sciences, Bartycka 18A, 00-716, Warsaw, Poland

[3]Southwest Research Institute, P.O. Drawer 28510, San Antonio, TX 78228, USA

[4]Los Alamos National Laboratory, Intelligence and Space Research Division, P.O. Box 1663, Los Alamos, NM 87545, USA

[5]University of Texas at San Antonio, San Antonio, TX 78249, USA

[6]University of Montana, 32 Campus Drive, Missoula, MT 59812, USA

[7]University of New Hampshire, Space Science Center, Morse Hall Rm 407, Durham, NH 03824, USA

[8]Solar-Terrestrial Environment Laboratory, Nagoya University, Nagoya 464-8601, Japan

## ABSTRACT

The Interstellar Boundary Explorer (IBEX) has now operated in space for seven years and returned nearly continuous observations that have led to scientific discoveries and reshaped our entire understanding of the outer heliosphere and its interaction with the local interstellar medium. Here we extend prior work, adding the 2014-2015 data for the first time and examine, validate, initially analyze, and provide a complete seven year set of Energetic Neutral Atom (ENA) observations from ~0.1 to 6 keV. The data, maps, and documentation provided here represent the tenth major release of IBEX data and include improvements to various prior corrections to provide the citable reference for the current version of IBEX data. We are now able to study time variations in the outer heliosphere and interstellar interaction over more than half a solar cycle. We find that the Ribbon has evolved differently than the globally distributed flux (GDF) with a leveling off and partial recovery of ENAs from the GDF, owing to solar wind output flattening and recovery. The Ribbon has now also lost its latitudinal ordering, which reflects the breakdown of solar minimum solar wind conditions, and exhibits a greater time delay than for the surrounding GDF. Together, the IBEX observations strongly support a secondary ENA source for the Ribbon and we suggest that this be adopted as the nominal explanation of the Ribbon going forward.

*Key words:* local interstellar matter – solar wind – Sun: activity – Sun: heliosphere – Sun: magnetic fields

# 1. INTRODUCTION

IBEX – the Interstellar Boundary Explorer (McComas et al. 2009a) – is a NASA mission that has been providing nearly continuous observations of the outer heliosphere and its interaction with the local interstellar medium (LISM) since the beginning of 2009 (see McComas et al. 2009c and other papers in the IBEX Special Issue of *Science*). These observations are unique in two ways: 1) IBEX provides the only global (all sky) measurements of hydrogen Energetic Neutral Atoms (ENAs) from the outer heliosphere, over the broad energy range from ~0.1 to 6 keV where the bulk of these emissions reside, and 2) IBEX has been providing these global measurements nearly continuously for more than half a solar cycle. The IBEX payload comprises two single pixel imagers: IBEX-Hi (Funsten et al. 2009a) covering energies from ~0.5 to 6 keV in six energy bins, and IBEX-Lo (Fuselier et al. 2009b) measuring from ~0.01 to 2 keV in eight energy bins. The IBEX spacecraft spins at ~4 RPM and its spin axis is repointed a few degrees west of the Sun every few days as its inertially-fixed direction drifts across the Sun at ~1° per day owing to Earth's orbital motion. This combination produces complete new sets of energy-resolved all-sky ENA maps every six months.

IBEX observations have led to numerous "firsts" and discoveries, including the first measurements of the globally distributed flux (GDF) of ENAs from the inner heliosheath and discovery of the completely unpredicted "Ribbon" of enhanced ENA emissions. These, and other important results were published in a special issue of *Science* in November 2009 (McComas et al. 2009c; Fuselier et al. 2009a; Funsten et al. 2009b; Möbius et al. 2009; Schwadron et al. 2009). These and numerous other firsts and discoveries from 2009 to 2013 were enumerated in Table 1 of McComas et al. (2014a), which also provided the first five full years of IBEX observations and examined temporal variations over that interval. Since the end of 2013, IBEX has extended the string of groundbreaking research and results and Table 1 of this study summarizes the major firsts and discoveries of the IBEX mission into 2016.

*Table 1. Major "firsts" and discoveries over IBEX's first seven years.*

| **IBEX Ribbon** | |
|---|---|
| Discovery of an enhanced ENA Ribbon flux and its connection to the interstellar magnetic field (See review by McComas et al. 2014a) | McComas et al. 2009c<br>Fuselier et al. 2009a<br>Funsten et al. 2009b<br>Schwadron et al. 2009 |
| Discovery of solar wind-like latitude/energy ordering of Ribbon emissions | McComas et al. 2012c |
| Discovery of different time variations in different portions of the Ribbon, consistent with latitude structure and secondary ENA Ribbon source | McComas et al. 2014a |
| Discovery of time variations in Ribbon at least down to 6 month scales | McComas et al. 2010 |
| Triangulation of interstellar magnetic field unfolding in Voyager 1 obs., and IBEX measurements of Ribbon and ISN flow | Schwadron et al. 2015b |
| Distance to Ribbon source inferred from parallax is 140 +84/-38 AU, consistent with secondary ENA source | Swaczyna et al. 2016a |
| The energy-dependent position of the IBEX Ribbon due to the solar wind structure | Swaczyna et al. 2016b |
| **Global Heliosphere** | |
| First observations of globally distributed flux (GDF) ENAs from the inner heliosheath | McComas et al. 2009c<br>Schwadron et al. 2009 |
| Discovery of rapid (~6 months) time variations in the heliosphere's interstellar interaction and connection to decreasing solar wind output | McComas et al. 2010<br>Reisenfeld et al. 2012, 2016<br>McComas et al. 2012c |
| GDF ENAs ordered by the latitudinal solar wind structure | Dayeh et al. 2011<br>Desai et al. 2015 |
| First observations of the heliotail, it's ordering by fast and slow solar wind, and the influence of the interstellar magnetic field | Schwadron et al. 2011a<br>McComas et al. 2013b |
| Estimates of partitioning of energy between termination shock-processed particle populations in the Voyager directions | Wu et al. 2010<br>Desai et al. 2014<br>Zank et al. 2010<br>Zirnstein et al. 2014 |
| Discovery of the thermodynamic state of inner heliosheath far from equilibrium | Livadiotis et al. 2011<br>Livadiotis & McComas 2013 |
| Discovery of region of maximum pressure in inner heliosheath and explanation of flow direction in inner heliosheath observed by Voyager 2 | Schwadron et al. 2014c<br>McComas & Schwadron 2014 |
| Discovery of the "flattening out" of the ENA fluxes ~2012-2013 over most of the sky other than the heliotail, where they continued to drop off | McComas et al. 2014a |
| Energy and latitude-dependence of ENA spectral indices indicates spectra not representable by single power law | Desai et al. 2015<br>Dayeh et al. 2012 |
| Asymmetry of heliosheath pressure and plasma flows, and connection of heliotail port/starboard lobes to TS/IHS geometry, PUI abundance, and IHS plasma flow properties | Schwadron et al. 2014c<br>McComas & Schwadron 2014<br>Zirnstein et al. 2016a |
| Roll-over of heliospheric neutral H below 100 eV | Fuselier et al. 2014<br>Galli et al. 2016 |
| Heliosheath pressure from the poles correlates with 11 year solar cycle | Reisenfeld et al. 2016 |
| 11 yr solar cycle and energy-dependent extinction of IHS PUIs by charge-exchange likely responsible for generating heliotail lobe structure | Zirnstein et al. 2016c |
| Voyager 1 in situ data outside the HP are consistent with a plasma depletion layer ~5 AU thick and connected to the B-V plane of the heliosphere based on IBEX observations | Cairns & Fuselier 2016 |

## Interstellar Medium

| | |
|---|---|
| First direct observations of interstellar Hydrogen, Deuterium, Oxygen, and Neon | Möbius et al. 2009 Bochsler et al. 2012 Rodríguez Moreno et al. 2013 |
| Discovery of secondary population of He (the "warm breeze") | Kubiak et al. 2014 |
| First connection of LISM environment from IBEX to TeV cosmic rays | Schwadron et al. 2014a |
| Discovery that the heliosphere might have a bow wave ahead of it instead of a bow shock | McComas et al. 2012a Zank et al. 2013 |
| First precise estimate of interstellar field strength as well as direction | Zirnstein et al. 2016b |
| Refined ISN He flow direction, temperature, and speed | McComas et al. 2015b Schwadron et al. 2015a Bzowski et al. 2015 |
| VLISM is warmer than previously expected | McComas et al. 2015a Möbius et al. 2015b |
| Co-planarity of ISN He, H, He Warm Breeze, the IBEX Ribbon center, and the interstellar magnetic field deduced from the Ribbon | ApJ Supp. Series 2015 Kubiak et al. 2016 Zirnstein et al. 2016b |
| Determination of the local gas Ne/O ratio from neutral flow observations | Bochsler et al. 2012 Park et al. 2014 |
| Confirmation of He and O secondary component possibly from the OHS | Park et al. 2015 |
| First quantitative derivation of ISN O properties, evidence for significant processing in OHS | Schwadron et al. 2016a |
| First direct sampling of ISN H and its evolution during the solar cycle | Saul et al. 2012, 2013 |
| Independent derivation of solar radiation pressure from ISN H observations revealed to be greater than that inferred from solar Ly-alpha flux data | Katushkina et al. 2015 |
| Possible IS dust filament in OHS and correlation with LISM inflow direction | Frisch et al. 2015 |
| First derivation of IS flow longitude from symmetry of IS PUI cut-off at 1AU and connection to IBEX measurements | Moebius et al. 2015c |

## Terrestrial Magnetosphere

| | |
|---|---|
| First imaging of Earth's subsolar magnetopause | Fuselier et al. 2010 |
| First imaging of dynamic magnetotail and possible disconnection event | McComas et al. 2011b |
| First images of magnetospheric cusps and their asymmetry | Petrinec et al. 2011 |
| First characterization of dayside magnetosheath using ENAs | Ogasawara et al. 2013 |
| First combined mission ENA imaging to provide direct timing of plasma transfer from dayside compression to magnetospheric ring current | McComas et al. 2012b |
| Motion of terrestrial plasma sheet dominated by seasonal and diurnal motion of Earth's dipole tilt | Dayeh et al. 2015 |
| First imaging of development of cold terrestrial plasma sheet during period of northward IMF and its reversal | Fuselier et al. 2015 |
| Evidence for suprathermal ion acceleration by diffusive shock acceleration at Earth's bow shock, shocked SW in subsolar magnetopause | Ogasawara et al. 2015 |
| **Moon** | |
| First measurement of neutralized and backscattered solar wind from the Moon | McComas et al. 2009b |
| Discovery of lunar ENA albedo on solar wind speed and Mach number | Funsten et al. 2013a Allegrini et al. 2013 |
| **Space Mission Capabilities** | |
| First use of additional Solid Rocket Motor on Pegasus LV and spacecraft propulsion to achieve very high altitude orbit | McComas et al. 2009a Scherrer et al. 2009 |
| Discovery and first use of long-term stable lunar synchronous orbit. | McComas et al. 2011a |

The IBEX Ribbon is a narrow (~20º wide from 0.7-2.7 keV; Fuselier et al. 2009a), nearly circular feature (Funsten et al. 2009b, 2013b) with ENA fluxes reaching ~2-3 times that of the surrounding GDF (McComas et al. 2009c). Its directional location appears to be ordered by the external magnetic field in the very local interstellar medium (McComas et al. 2009c; Schwadron et al. 2009). Simultaneously, the dominant ENA emissions reflect the latitude-dependent energy distribution of the out-flowing solar wind (McComas et al. 2012c) over the past, protracted solar minimum (McComas et al. 2008; 2013a), indicating that the solar wind must be the ultimate source of the Ribbon ENAs.

There are over a dozen different ideas, models, and scenarios for how the Ribbon could be generated (see review papers McComas et al. 2011b, 2014b; and new models by Isenberg 2014; Giacalone & Jokipii 2015). The leading candidates are various versions of a "secondary ENA" process, derived from a multistep interaction where 1) some fraction of the solar wind and inner heliosheath ions are neutralized and radiate outward, 2) these "primary" neutrals are re-ionized and gyrate around the interstellar magnetic field of the very local interstellar medium just beyond the heliopause (the outer heliosheath), and 3) eventually these ions charge exchange again and produce secondary ENAs that preferentially radiate back inward toward the Sun from regions where the draped interstellar magnetic field is perpendicular to a radial line of sight from the Sun (and IBEX) (McComas et al. 2009c). Various detailed models and calculations produce very Ribbon-like ENA fluxes even though they are based on different ideas for how to get the secondary ENAs to preferentially propagate back inward from regions where the field is perpendicular to the radial: perhaps the gyrating ions stay in ring-beam distributions for the several years required to re-neutralize (Heerikhuisen et al. 2010; Chalov et al. 2010; Möbius et al. 2013), or in the opposite case of strong scattering and wave-particle interactions, these processes may cause the ions to be spatially confined (Schwadron & McComas 2013; Isenberg 2014). The spatial confinement may be associated with pre-existing turbulence causing magnitude fluctuations in the local interstellar magnetic field sufficient to trap pitch angle distributions owing to magnetic mirroring (Giacalone & Jokipii, 2015), or a combination of large-scale and small-scale turbulence that produces a marginally-stable ion distribution (Gamayunov et al. 2010). Zirnstein et al. (2015a) recently provided a summary table of Ribbon observables reproduced by various secondary ENA mechanisms.

As the IBEX database has grown, we have been able to study time variations in the ENA fluxes arriving from the outer heliosphere. These prior studies of the all-sky variations included just the first year (McComas et al. 2010), the first three years (McComas et al. 2012c), and first five years of IBEX observations (McComas et al. 2014a). Five other studies focused on time variations in the ENA fluxes from the polar regions (Reisenfeld et al. 2012, 2016; Allegrini et al. 2012; Dayeh et al. 2012, 2014); these are especially critical as IBEX's viewing and sampling provide both essentially continuously measurements of these directions and better statistics.

In principle, measurements in the polar directions could reveal time variations even faster than the six months "revisit" time that IBEX has for the rest of the sky. However, all studies to date have shown time variations no faster than roughly six months, which is consistent with variations in the global solar wind output, including a general reduction in the ENA fluxes, consistent with the long-term decreasing solar wind output (McComas et al. 2008; 2013a). The most complete all-sky study so far (McComas et al. 2014a) showed time variations of ENA emissions from both IBEX Ribbon and GDF, with both decreasing from 2009 to 2011, and then evidence for stabilization and even some recovery of fluxes from ~2011 to 2013. Moreover, Reisenfeld et al. (2016) revealed the evolution of heliosheath pressure from the poles are consistent with the 11-year solar cycle and the closing of the polar coronal holes.

This study extends our prior work (McComas et al. 2012c, 2014a) and provides the documentation for the release of the sixth and seventh years (2014 and 2015) of IBEX data, as well as the re-release, with slightly improved background subtraction and correction factors, of years one through five (2009-2013). Section 2 shows the seven years of data, largely following the format of our prior studies. We examine time variations of the ENA fluxes observed by IBEX over more than half a solar cycle of data in Section 3 and in Section 4 we discuss the implications of these new observations for our understanding of the heliosphere's interaction with the local interstellar medium. As in our prior studies, the appendices provide additional detailed documentation useful to outside researchers using the IBEX data. Appendix A provides a listing of the specific source files at the ISOC used to generate the figures shown in this study, while Appendix B follows the methodology introduced by McComas et al. (2012c, 2014a) and provides updated orbit-by-orbit survival probability corrections for both IBEX-Hi and -Lo data used in this paper.

Thus, this study provides the citable reference for the first seven years of IBEX data and for the corrections to and validation of the best possible data set that the IBEX team can currently provide. As with all prior IBEX data releases, these data are available at: http://ibex.swri.edu/researchers/publicdata.shtml, through the data section of the general IBEX web site: http://ibex.swri.edu/, and in the archive at the National Space Science Data Center (NSSDC): http://nssdc.gsfc.nasa.gov/.

## 2. SEVEN YEARS of IBEX OBSERVATIONS

The IBEX spacecraft is a Sun-pointed spinner (~4 RPM), with two single pixel ENA cameras that view perpendicular to the spin axis (McComas et al. 2009a). This configuration means that for each spacecraft rotation, IBEX samples ENAs from a predetermined (great circle) band around the sky. Then, as the spacecraft is periodically repointed to maintain its nearly sun-pointed configuration, adjacent bands of the sky are viewed such that complete sky viewing (and the production of full sky maps) is achieved every half year. The health of the spacecraft and instruments – IBEX-Hi (Funsten et al. 2009a) and IBEX-Lo (Fuselier et al. 2009b) – remains excellent and IBEX has now made nearly continuous observations of ENAs from the outer heliosphere for over seven years.

While the basic observational strategy has been the same throughout, a significant operational change was made in June of 2011. At that time, the IBEX team maneuvered the spacecraft into a long-term stable lunar synchronous orbit (McComas et al. 2011a), where it will remain beyond 2050 (longer, but we cut off the orbital calculations at this point). Prior to the maneuver, IBEX's orbital period was ~7.5 days, while after it was increased to ~9.1 days. For the shorter period, we only repointed the spacecraft once per orbit, around perigee. After the maneuver (orbit 130) we started repointing twice per orbit, around both perigee and apogee. Thus, while data from full orbits were combined before orbit 130, producing viewing bands offset by ~7.5 °, thereafter, data are combined separately for the ascending (designated "a") and descending ("b") portions of each orbit providing observational viewing bands offset by ~4.5°.

Table 2 provides the dates and orbit/orbit arc numbers for all 14 energy-resolved sets of six-month and seven full year maps. For this study, we have improved the naming convention from what we previously called "odd" and "even" to the maps corresponding to roughly the first six months of each year (A maps) and second six months of each year (B maps); this convention makes it easier to immediately identify when each map's data were taken. Thus, what would previously have been called the odd maps (1, 3, 5, 7, 9, 11, 13) are now 2009A-2015A and the even maps (2, 4, 6, 8, 10, 12, 14) are now 2009B-2015B.

*Table 2. Data intervals used for the first seven years of IBEX maps; years 1-5 are unchanged from McComas et al. (2014a) while years 6-7 are new.*

| Year (Annual Maps) | 6-month Maps | Orbit/Arc Numbers | Dates (start/end of orbits or arcs) |
|---|---|---|---|
| Year 1 (2009 Map) | 1 (2009A) | 11-34 | 12/25/2008 – 06/25/2009 |
| | 2 (2009B) | 35-58 | 06/25/2009 – 12/25/2009 |
| Year 2 | 3 (2010A) | 59-82 | 12/25/2009 – 06/26/2010 |

| (2010 Map) | 4 (2010B) | 83-106 | 06/26/2010 – 12/26/2010 |
|---|---|---|---|
| Year 3 | 5 (2011A) | 107-130a | 12/26/2010 – 06/25/2011 |
| (2011 Map) | 6 (2011B) | 130b-150a | 06/25/2011 – 12/24/2011 |
| Year 4 | 7 (2012A) | 150b-170a | 12/24/2011 – 06/22/2012 |
| (2012 Map) | 8 (2012B) | 170b-190b | 06/22/2012 – 12/26/2012 |
| Year 5 | 9 (2013A) | 191a-210b | 12/26/2012 – 06/26/2013 |
| (2013 Map) | 10 (2013B) | 211a-230b | 06/26/2013 – 12/26/2013 |
| Year 6 | 11 (2014A) | 231a - 250b | 12/26/2013 – 6/26/2014 |
| (2014 Map) | 12 (2014B) | 251a - 270b | 6/26/2014 – 12/24/2014 |
| Year 7 | 13 (2015A) | 271a - 290b | 12/24/2014 – 6/24/2015 |
| (2015 Map) | 14 (2015B) | 291a - 310b | 6/24/2015 – 12/24/2015 |

In orbit segment 184a, we made an additional change, modifying the IBEX-Hi energy step sequence from ESA 1-2-3-4-5-6 to 2-3-3-4-5-6; this change removed ESA 1, which was often noisy and doubled the acquisition time for ESA 3 (center energy ~1.1 keV), where the Ribbon is most easily observed. Energy ranges of the passbands for ESAs 2-6 are given in Table 3 of McComas et al. (2014a).

The first 19 figures in this study show various sets of IBEX sky maps and other plots as in the five-year paper (McComas et al. 2014a). For ease of comparison, we have chosen to provide them in the same order and with each having the same figure number as the equivalent figures in the two papers. To accommodate the additional two years of data, we transposed the display format so that now each column provides data from a particular energy passband and the rows represent observations from different times (six-month individual maps and 12-month annual maps), with the earliest observations at the top and progressively later observations down the page.

## 2.1 IBEX ENA Data Processing

As in the 3- and 5-year papers (McComas et al. 2012c, 2014a), we use the most reliable triple coincidence events (Funsten et al. 2009a) to produce IBEX-Hi flux maps. Also, as we did in those studies, we first "cull" out all times of enhanced backgrounds. These include: 1) whenever there are high count rates in the IBEX Background Monitor (Allegrini et al. 2009); 2) whenever there are enhanced counts at lower energies over a broad range of spin-phases; 3) whenever the Moon or the Earth's magnetosphere is in the field of view; 4) whenever there are enhanced solar energetic particles (SEPs); and 5) very rare bursts of counts generated internally to the instrument. The data set used in and being released with this study includes years 6 and 7 and provides slightly improved culling for several orbits in years 1-5.

Corrections for always-present backgrounds are applied in the same manner as in the 5-year paper (McComas et al. 2014a). The data here include a slightly improved correction for the time-variable cosmic ray background for the first five years as well as its extension through years 6 and 7.  We also include slightly improved corrections for a residual background produced by the "ion gun" effect inside IBEX-Hi, generated by acceleration of ions produced by electron impact ionization of residual neutral atoms and molecules within IBEX-Hi (this background was considerably reduced in data collected since mid-2013 by optimization of collimator voltages).  Finally, as done in McComas et al. (2012c, 2014a), we include times after subtracting a small additional isotropic background for some orbits where statistics are low; this process improves the statistical accuracy of otherwise poorly resolved swaths. We also incorporated new orbit-by-orbit survival probability corrections for orbits in the first five years of observations as well as years 6 and 7 (Appendix B). The most significant change in data processing since the 5-year paper is the application of a time-varying efficiency correction to the IBEX-Hi ENA count rates.  The ENA detection efficiency can be determined in situ for two of the three IBEX-Hi detector sections, based on ratios of count rates for various coincidence types after penetrating background counts for these coincidence types have been subtracted (McComas et al. 2014a, Appendix C).  The accuracy of these efficiency determinations increases as the overall statistics increase.

After 5 years, there was some suggestion that detector section efficiency may have decreased slightly. Periodic gain tests also indicated a gradual increase in the voltage of the lower edge of the gain plateau for the CEMs in the IBEX-Hi detector, bringing the edge closer to (but still below) the operating voltage.  In the first half of 2014 (2014A), the CEMs were run at their original voltage (1700V) and at a slightly increased voltage (1780V), alternating between these two levels twice per orbit arc.  This special process allowed a precise relative calibration of IBEX-Hi between the two operational high voltage levels.  Following the first half of 2014, the CEMs have been operated only at the increased voltage during collection of science data.  In this paper, for the first half of 2014, only data taken with the CEMs at 1780V have been incorporated into the maps.

Analysis of coincidence data by the method detailed in the 5-year paper (McComas et al. 2014a, Appendix C) with improved time-variable cosmic ray background subtraction, results in the conclusion that IBEX-Hi detector-section triple-coincidence efficiency dropped by roughly 10% linearly over the first year, whereupon it stabilized.  The increase in CEM operating voltage to 1780V increased the detector-section efficiency by approximately 6%.  These efficiency changes are included in the fluxes reported in this study.

IBEX-Lo data used in this study include IBEX-Lo's top four energy passbands (5-8) (Fuselier et al. 2009b), which cover the range from ~0.15 to 2.6 keV (FWHM). The culling procedure for IBEX-Lo year 4 and 5 data is similar to that of IBEX-Hi and generally the same as used previously in our three and five-year studies (McComas et al. 2012c, 2014a). Details of the IBEX-Lo map processing are described in Fuselier et al. (2012). In addition, as done by McComas et al. (2014a), we remove an additional background produced inside IBEX-Lo by the sputtering of neutrals (McComas et al. 2014a, Appendix E).

New for this study is the subtraction of a ubiquitous background in IBEX-Lo's energy pass bands 5 and 6. Galli et al (2014) conducted a detailed investigation of this background. They found that this local background appears to be associated with the near-Earth environment, the local environment around the spacecraft, or internal to the instrument. The background level is independent of time and look direction but depends on energy. The highest two energy passbands are unaffected, but there is a relatively small effect in energy passband 6 (~10% of the average signal) and a much larger effect in energy passband 5 (>50% of the average signal).

Finally, in this study (Appendix A) we provide a listing of the specific source files at the IBEX Science Operations Center (ISOC) used to generate each of the data figures in this study; these will make it much easier for other researchers to reproduce figures presented here if they so wish.

## 2.2 IBEX-Hi Maps in the Spacecraft Frame

As a standard product, the IBEX team generally displays sky maps of incoming ENAs in Mollweide projections. For most purposes, these are centered on the relative direction of the incoming interstellar flow into the heliosphere. IBEX has also contributed substantially to the direct observation of interstellar neutrals entering the heliosphere (e.g., Möbius et al. 2009; McComas et al. 2015b; Bzowski et al. 2015; Schwadron et al. 2015a). While the changes have been small, in a recent Astrophysical Journal Supplement Series, we derive a current best estimate of the inflow direction, with ecliptic longitude and latitude ($\lambda_{ISM\infty}$, $\beta_{ISM\infty}$) of (255.7°, 5.1°) (McComas et al. 2015b) (note: this is the inflow direction, whereas the relative LISM flow direction is opposite).

Figures 1 and 2 show IBEX-Hi all-sky maps of the observed ENA fluxes for the first half (A, aka "odd numbered maps") and second half (B, aka "even numbered maps") of each year, respectively. For consistency and ease of comparing different maps, we use the same color bars for each energy band across all the various figures throughout this study. Even in this simple display, it is easy to see the significant reduction in ENA fluxes from the IBEX Ribbon and essentially all parts of the sky over the first several years and a relative leveling off in the latter years at all energies in certain parts of the sky (see Section 3).

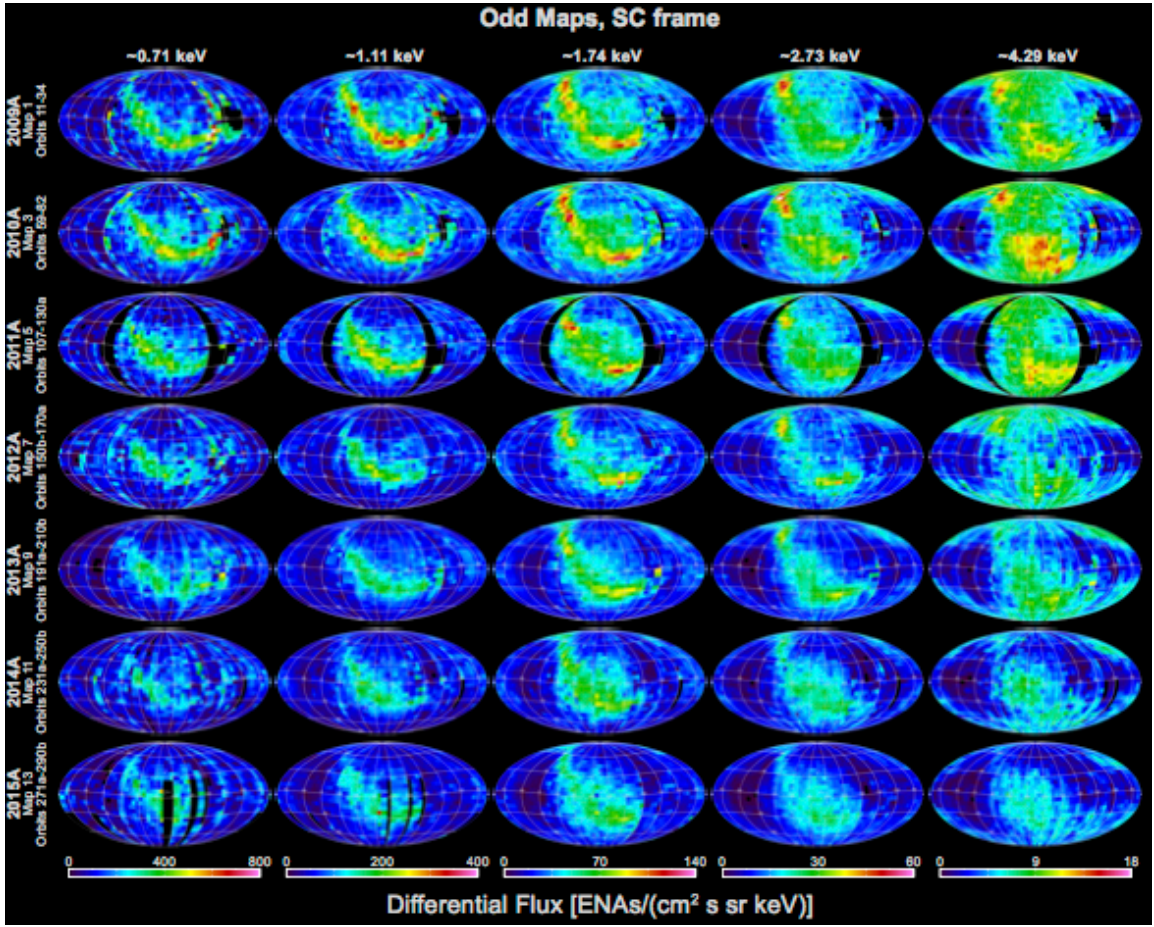

**Figure 1.** *Mollweide projections of IBEX-Hi ENA flux maps from the first half of each year (A maps). Each column represents a particular energy passband while the rows are from different years. Black regions indicate no data.*

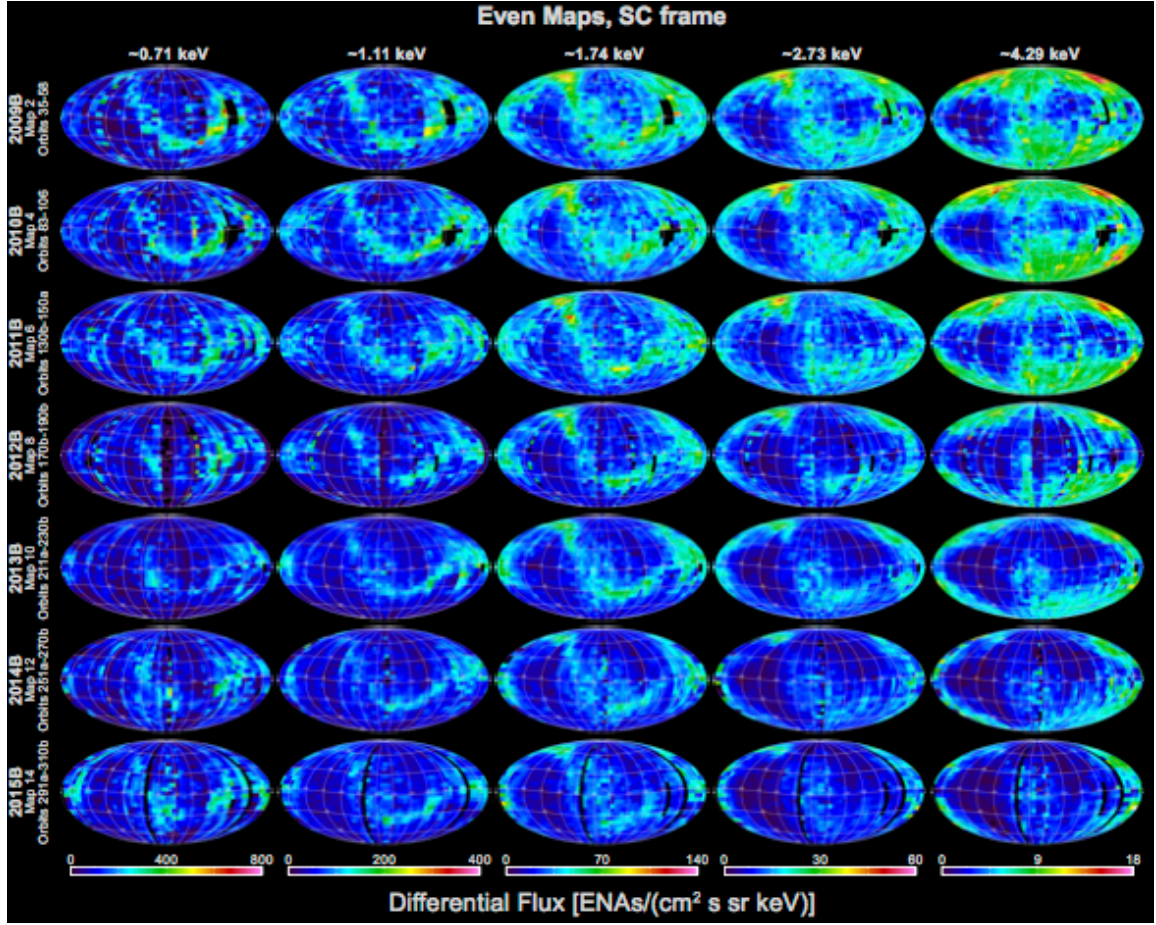

***Figure 2.*** *Same as for Figure 1, but for the second half of each year (B) map.*

## 2.3 IBEX-Hi Maps in the Inertial Frame

The significant differences between the fluxes shown in the first and second halves of the year maps are due to the motion of the spacecraft with respect to the incoming ENAs. This Compton-Getting (C-G) effect is largely due to the Earth's orbital motion about the Sun (~30 km s$^{-1}$). As shown explicitly in Figure 4 of McComas et al. (2012c), the C-G correction is a function of latitude and viewing angle with respect to the incoming ENAs. Thus, for the map intervals defined in Table 1, fluxes are enhanced across the central portion and reduced on the left and right sides of the A maps and reduced in the central portion and enhanced on the left and right sides of the B maps. The C-G effect also modifies the observed energy ranges in the various bands with lower intrinsic energies sampled in the ram viewing direction and higher energies on the anti-ram, particularly at the lowest energies and latitudes.

Using the improved procedure of McComas et al. (2012c, 2014a), we C-G correct the IBEX data in both energy and angle. Figures 3 and 4 show the C-G corrected data for the A and B maps, respectively. We note that C-G corrected maps need to be used carefully as the correction process can also introduce errors and artifacts.

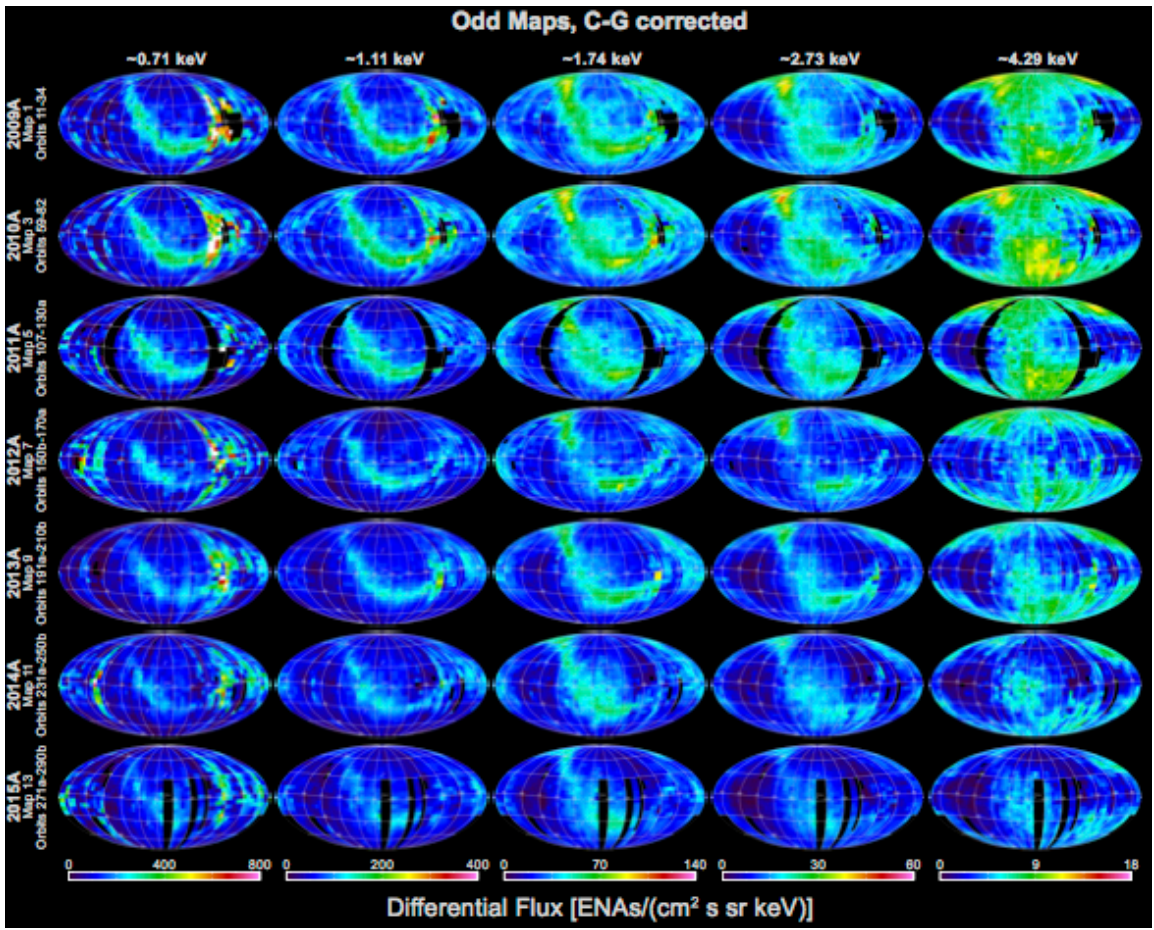

**Figure 3.** *IBEX ENA first half year (A) maps as in Figure 1, but C-G corrected into the heliospheric reference frame.*

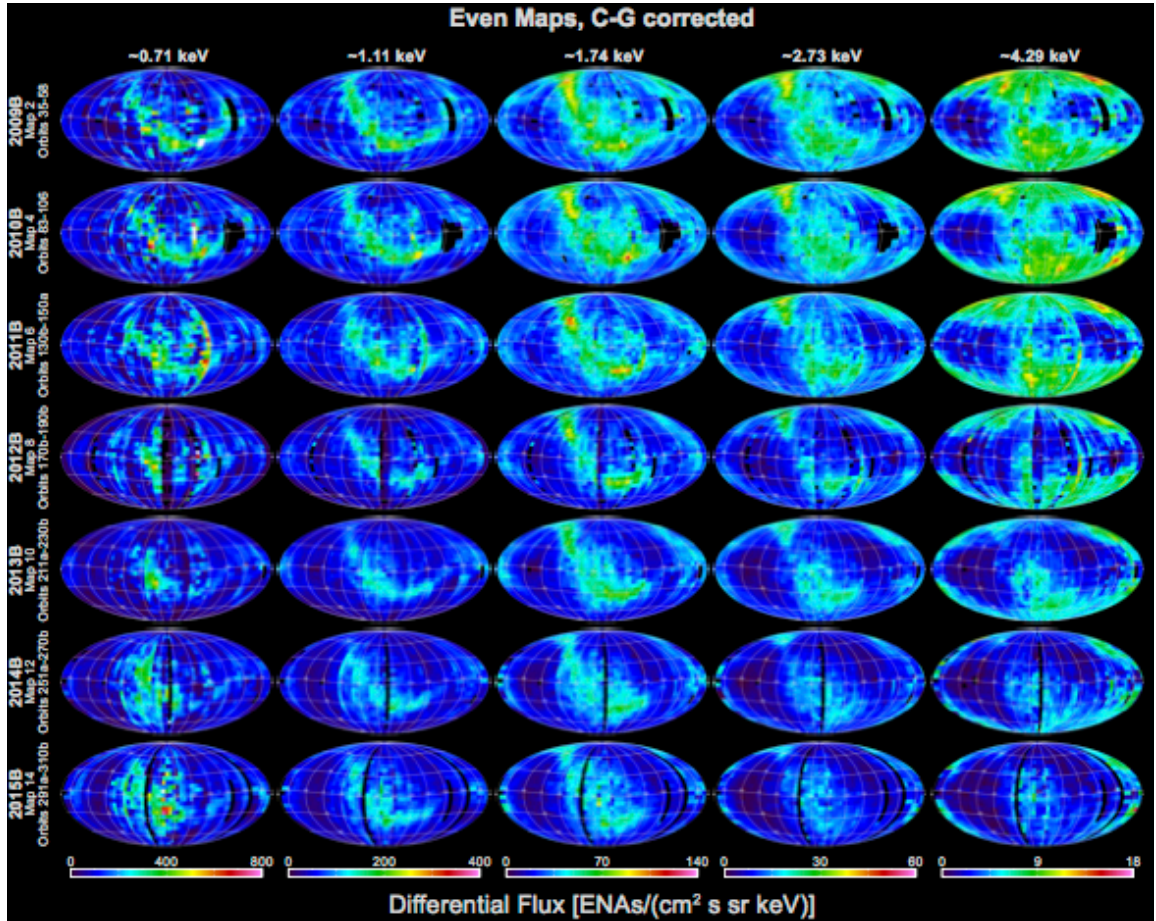

**Figure 4.** *Same as for Figure 3, but for the second half year (B) ENA maps.*

Figure 5 shows the combination of all seven years of C-G corrected IBEX data (2009-2015). As shown in McComas et al. (2010, 2012c, 2014a) and even more below, the ENA flux is somewhat variable over time, so statistically combining data from different times averages over these differences. Still, the maps shown in Figure 5 represent the "best" average ENA flux measurements observed at ~1 AU, in the heliospheric reference frame, over the 2009-2015 epoch.

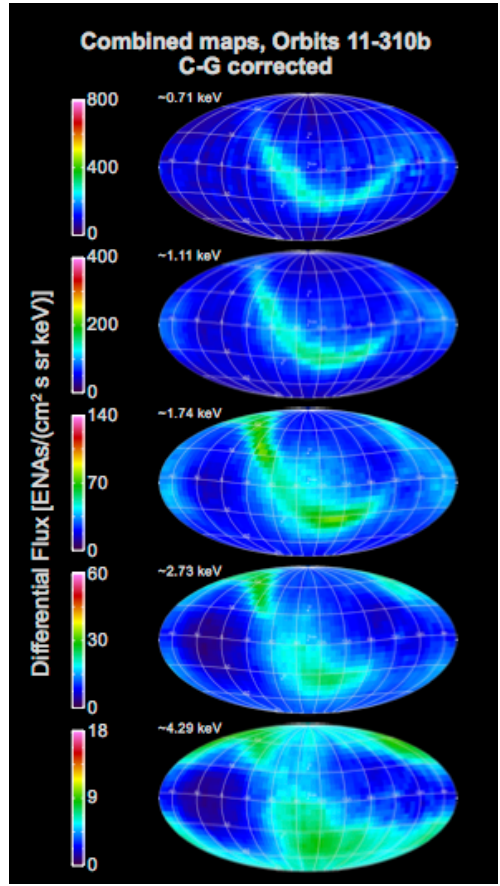

***Figure 5.*** *Combined ENA fluxes from all seven years of data in the heliospheric reference frame.*

## 2.4 IBEX Maps with Survival Probability Correction

The prior 3- and 5-year (McComas et al. 2012c, 2014a) studies employed a correction for the ENA flux observed directly at 1 AU by IBEX. In particular, we adjusted the fluxes to account for radiation pressure and ionization losses as ENAs transit inward from the outer heliosphere. The correction is both energy- and heliolatitude-dependent, and varies over time as the Sun and solar wind vary. The correction is relatively small beyond ~10 AU, but increases quickly as ENAs pierce within the innermost few AU of the heliosphere. Appendix B documents the orbit-by-orbit survival probabilities used in this study for both the IBEX-Hi and IBEX-Lo observations.

Figure 6 shows samples of the calculated survival probabilities for the northern (top) and southern (bottom) polar pixels for IBEX-Hi ESAs 2-6. The blue points show the survival probability calculated using available solar wind data (see Appendix B). Extrapolations beyond that point are shown by the red points. In the latest orbits, the survival probabilities have been decreasing in the northern polar pixel, but have started increasing in the southern one. Clearly these sorts of detailed orbit-by-orbit and pixel-by-pixel corrections are critical to inferring the correct source fluxes generated in the outer heliosphere and beyond.

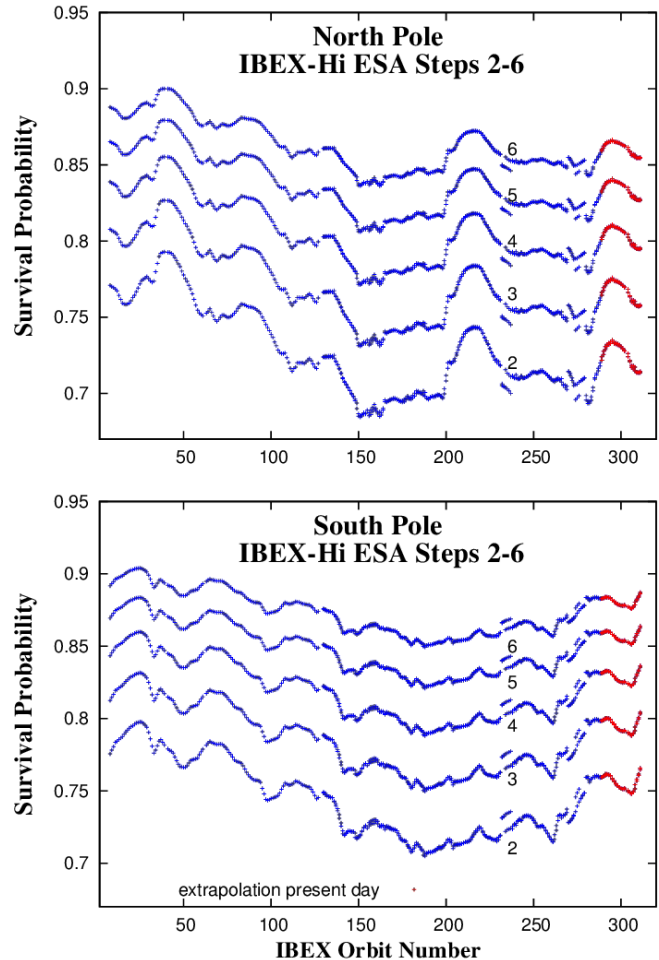

***Figure 6***. *Calculated survival probabilities for ENAs observed in IBEX's northern (top) and southern (bottom) polar pixels. Curves for the different ESA (energies) are indicated by the different number labels in each panel.*

Figures 7 and 8 show IBEX data including both survival probability and C-G corrections for the A and B maps, respectively. Similarly, Figure 9 shows the survival probability corrected combined maps for the 2009-2015 epoch. Just as Figure 5 shows the complete IBEX data as observed at 1 AU, Figure 9 provides the complete IBEX data on the flux of inward propagating ENAs around the vicinity of the termination shock, which is sunward of the region where they are generated. These data should be compared with theories and models of the sources of ENAs, which do not include their losses in transit in to 1 AU, and do not include time-dependent variations.

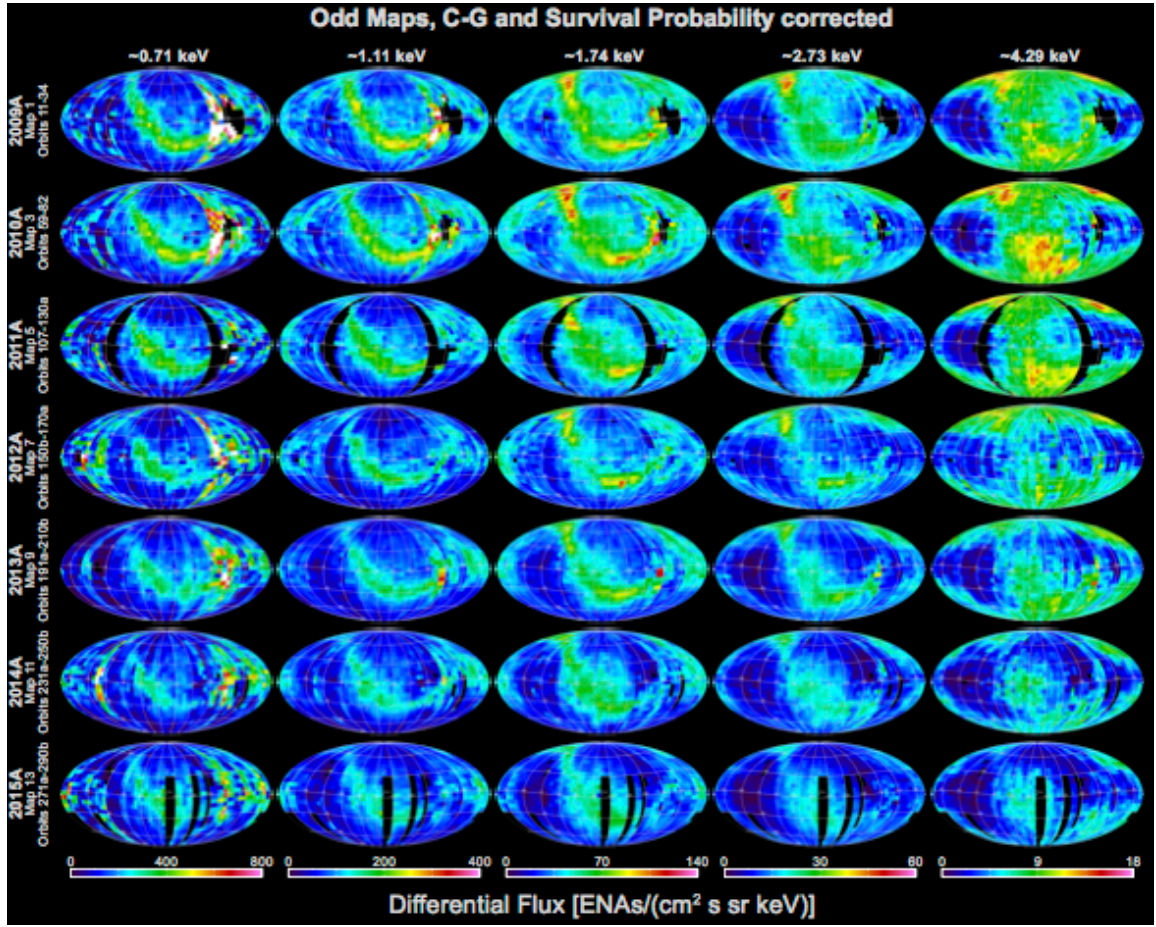

**Figure 7.** *First half year (A) ENA flux maps including survival probability and C-G corrections; these represent the expected inward-directed ENA fluxes around the termination shock.*

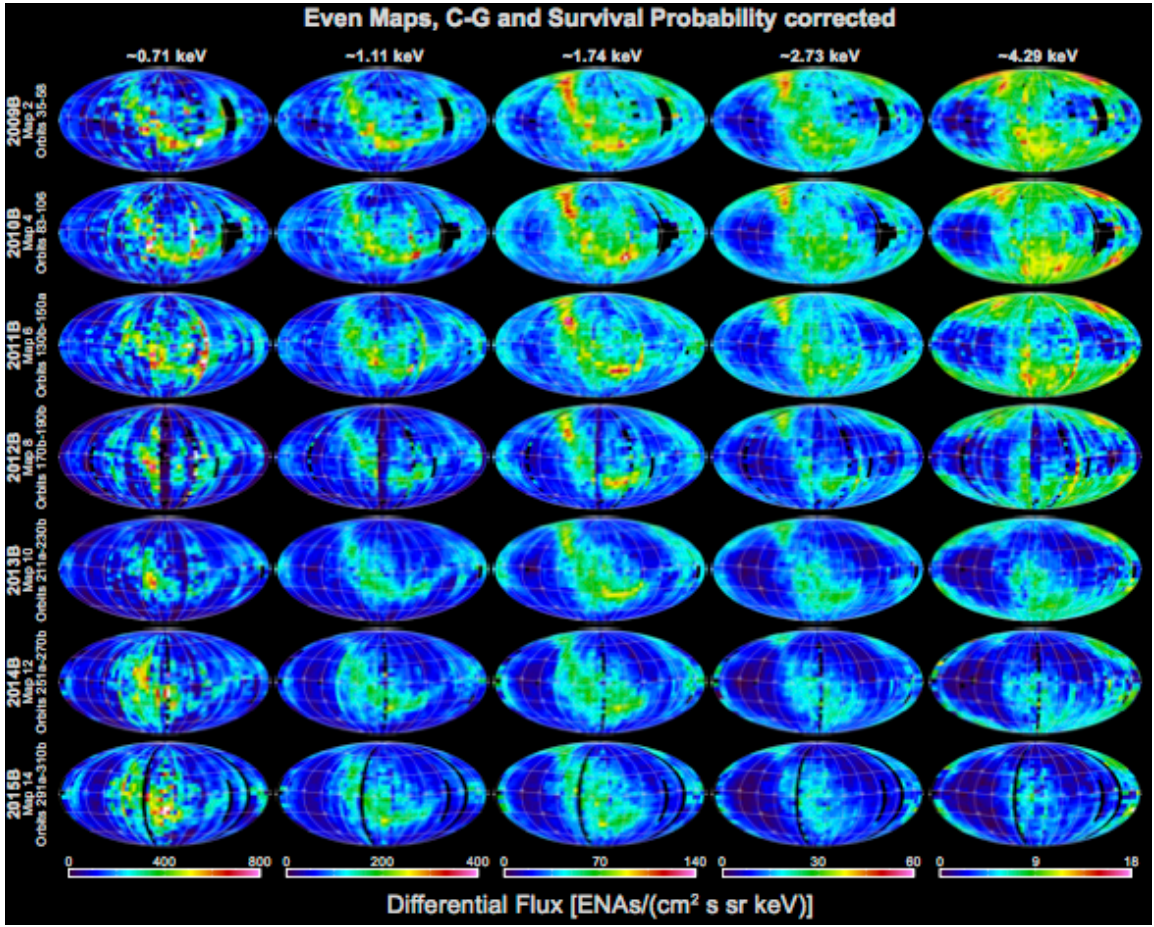

**Figure 8.** *Similar to Figure 7, but for second half year (B) maps.*

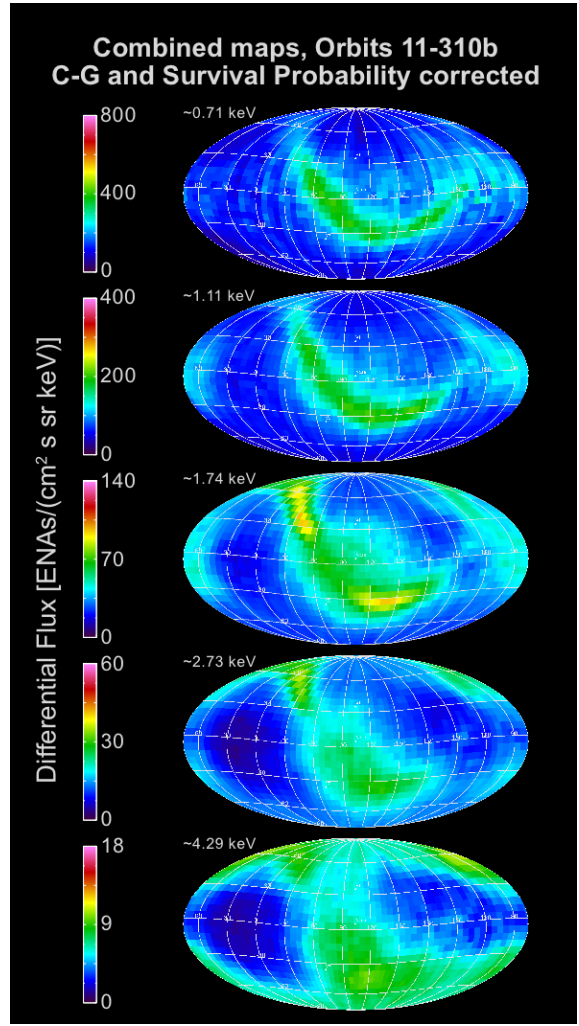

**Figure 9.** *Combined survival probability and C-G corrected maps, indicative of the average inward directed ENA fluxes around the termination shock over the 2009-2015 epoch.*

### 2.5 IBEX-Hi Ram and Anti-Ram Maps

Annual ram (anti-ram) maps (McComas et al. 2012c, 2014a) are derived from each year's worth of observations by combining all spin phases where the spacecraft's motion caused the sensors to be ramming into (retreating from) the ENAs. While different energies are still sampled at different latitudes, each pixel in the sky measures fluxes at the same energy from one year to the next. Thus, these maps are ideal for comparison between different years and for examining temporal variations without the uncertainties introduced by C-G corrections. On the other hand, because of the temporal changes in the solar output and solar wind conditions, the fluxes still need to be corrected for survival probability in transit from the outer heliosphere. Figures 10 and 11 provide these survival probability corrected ram and anti-ram maps for years 2009-2015. Figures 12 and 13 provide the statistically combined five-year maps for the ram and anti-ram directions, respectively.

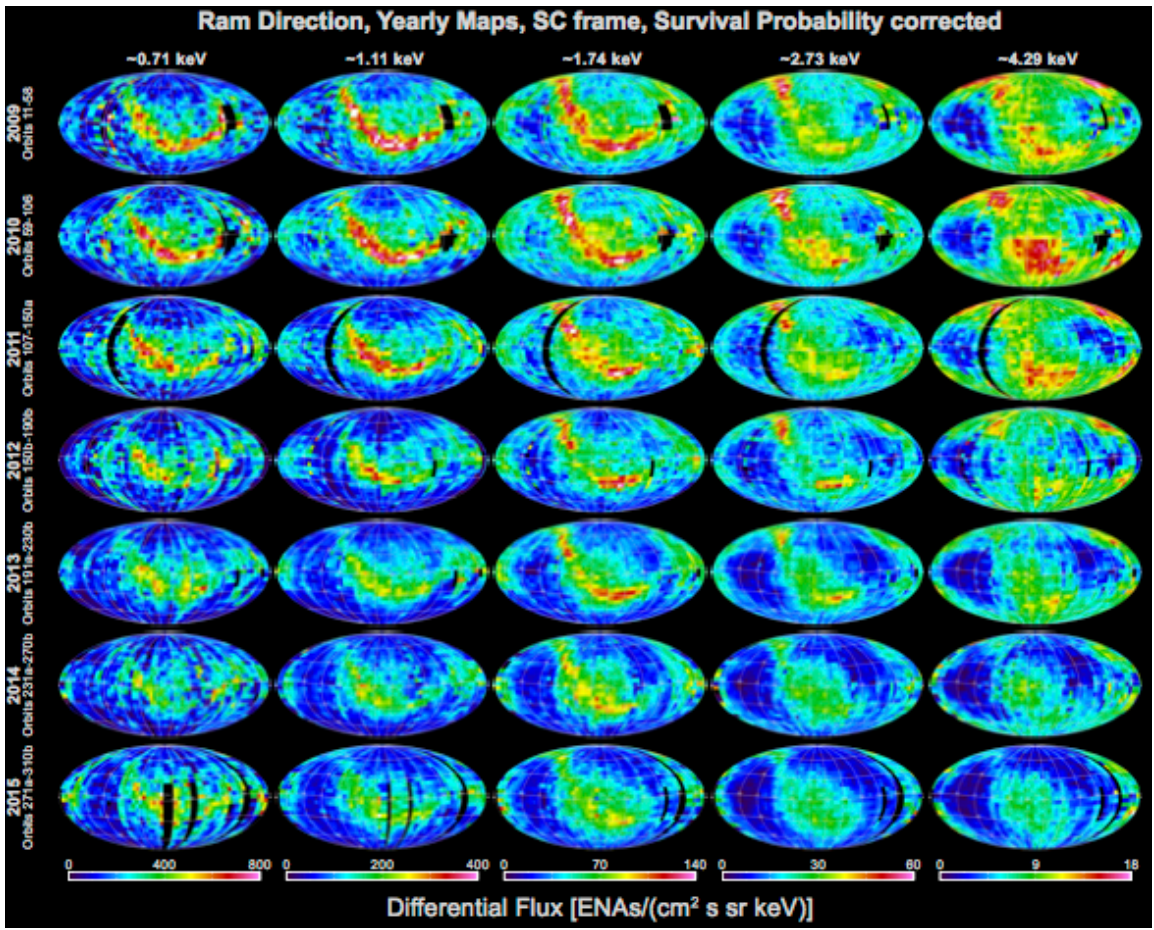

**Figure 10.** *Annual "ram" maps for 2009-2015 from IBEX-Hi data; fluxes are corrected for ENA survival probability.*

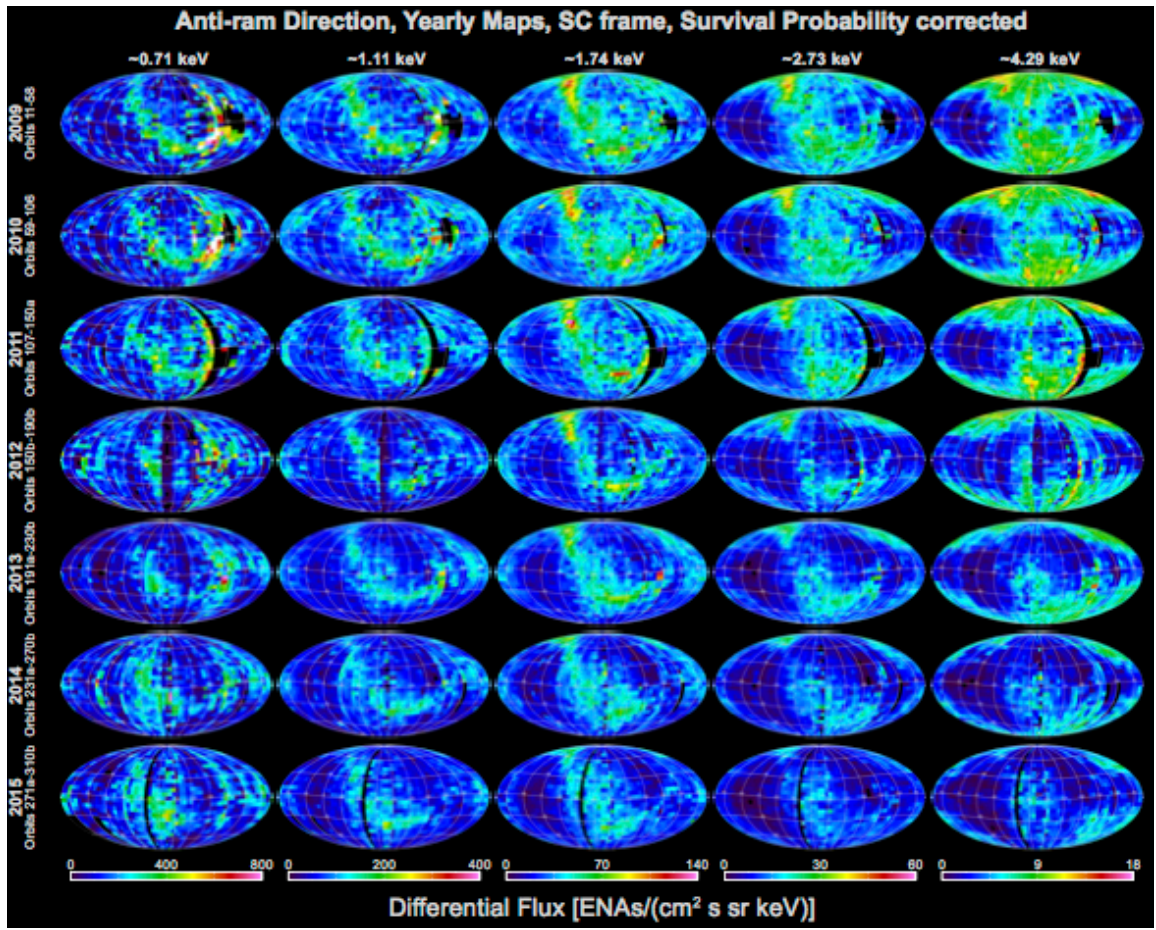

**Figure 11.** *Similar to Figure 10, but for "anti-ram" observations.*

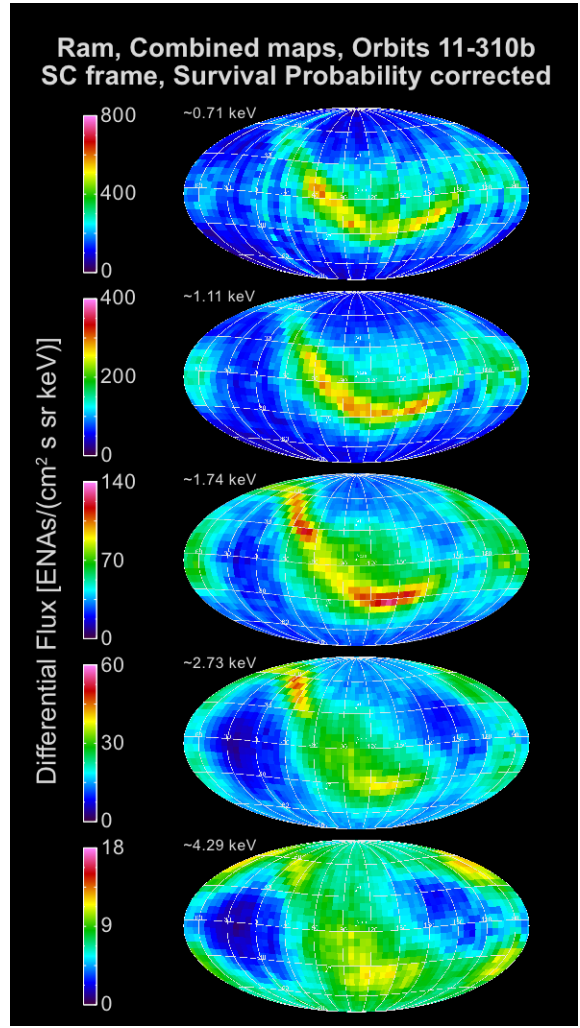

**Figure 12.** *Ram maps produced by statistically combining all seven annual ram maps at each energy from Figure 10.*

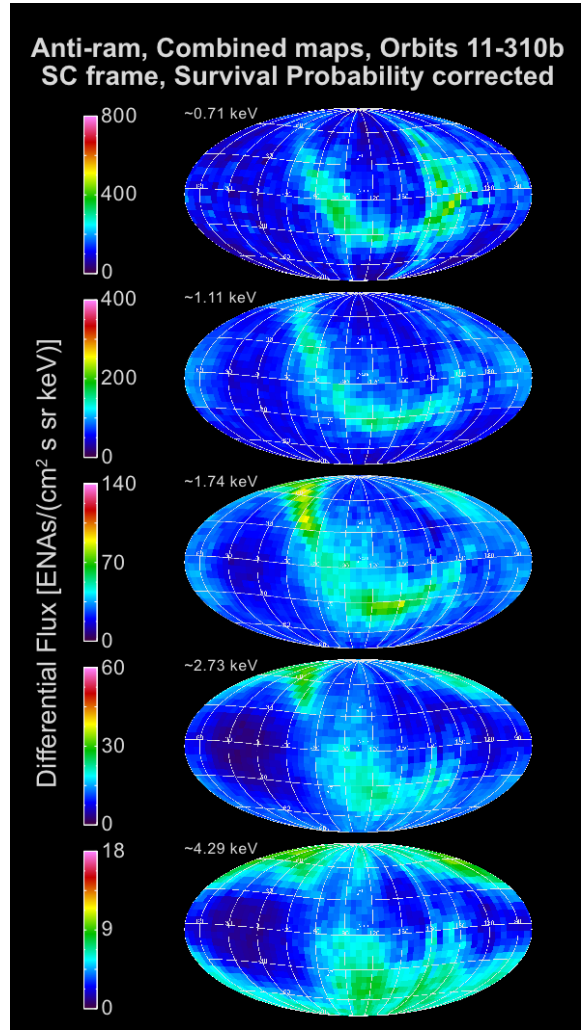

***Figure 13.*** *Similar to Figure 12, but anti-ram maps.*

## 2.6 IBEX-Lo Maps

As we did in the earlier three- and five-year studies, we also include IBEX-Lo maps for passbands with central energies of ~0.2, 0.4, 0.9, and 1.8 keV, which collectively cover energies from ~0.15 to 2.6 keV FWHM. Figure 14 provides maps combined over all seven years of observations and includes the correction to remove the signal from sputtering of ENAs and interstellar neutrals within IBEX-Lo (see Appendix E of McComas et al. 2014a) and the ubiquitous background in energy passbands centered at ~0.2 and 0.4 keV. In this figure, IBEX-Lo maps are displayed in the spacecraft frame and corrected for the survival probability, to account for the large losses of ENAs in transit from the outer heliosphere at these low energies. Comparison of the two columns of maps show that while nearly the entire sky has signal/noise (S/N) >3 at the highest energy, only the Ribbon and upwind direction, especially toward the starboard side at ~0.2 and 0.4 keV, have this statistical significance. The rest of the sky appears to have low fluxes from all directions, but we stress that these are at lower statistical significance, and thus, urge caution in using these lower energy ENA fluxes from other portions of the sky.

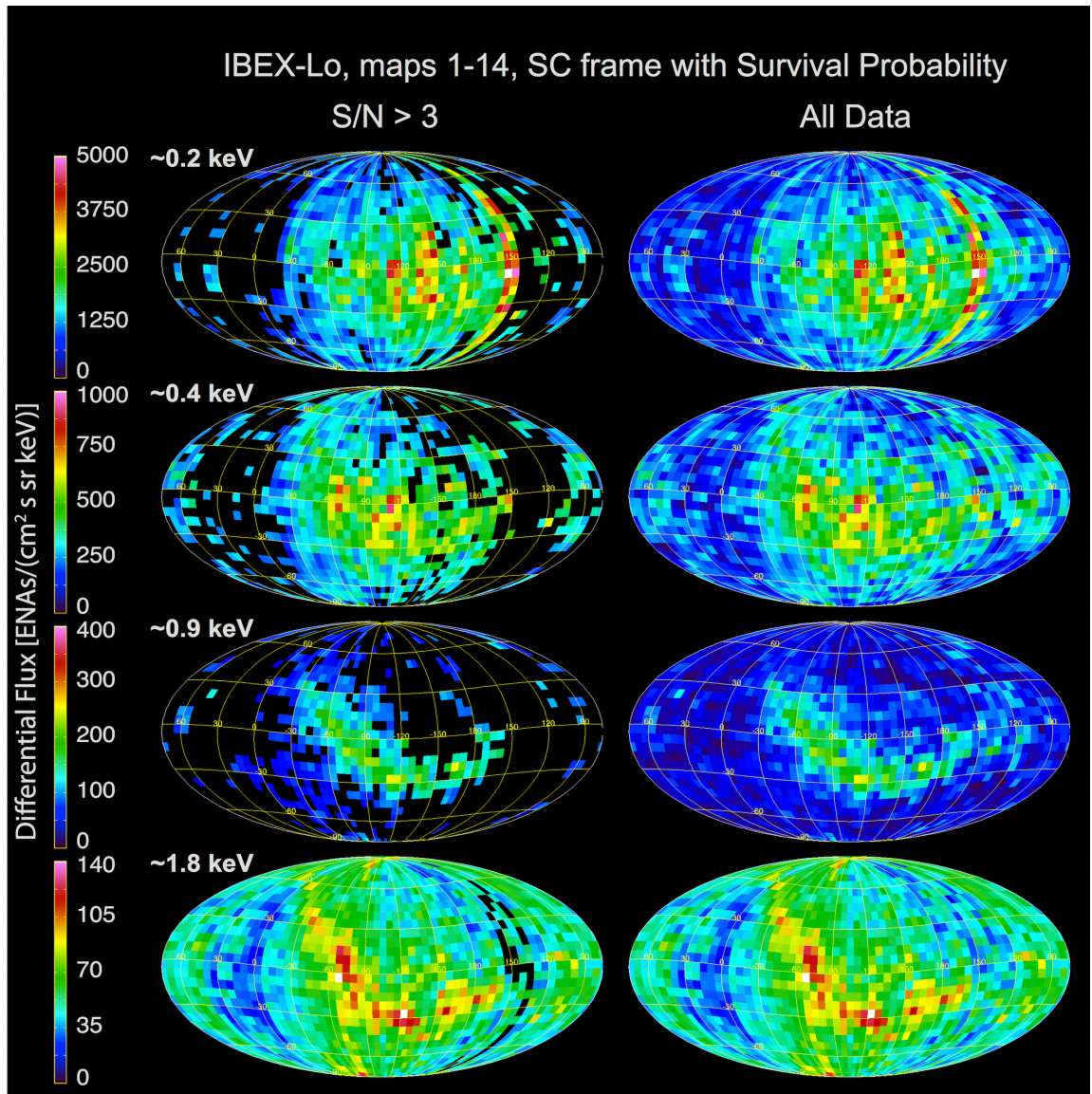

**Figure 14.** *Survival probability corrected IBEX-Lo maps in the spacecraft frame, including a S/N >3 (left) and no S/N (right) requirement. Energy passbands are centered at ~0.2, 0.4, 0.9, and 1.8 keV.*

Because of the totally unexpected discovery of the IBEX Ribbon, the higher energy IBEX-Lo observations, in the energy range overlapping the IBEX-Hi data, played an especially critical role. These provided a fully independent confirmation and validation of the IBEX-Hi observations of the Ribbon, required for a discovery of such a magnitude (McComas et al. 2009c). In Figure 14, the Ribbon is clear in the passbands centered at ~0.9 and 1.8 keV, consistent with IBEX-Hi observations.

While the IBEX-Lo data clearly have lower statistical significance than IBEX-Hi, they continue to be important, especially at the lower energies where IBEX-Lo provides the only measurements of heliospheric ENAs. With only the first three years of data, McComas et al. (2012c) showed that the Ribbon dimmed at the lower energies of ~0.2 and 0.4 keV, and McComas et al. (2014a) found that the Ribbon may be almost non-existent by the lowest energy step of ~0.2 keV (see also Galli et al. 2014). Similar to McComas et al. (2014a), we find here that the ~0.2 keV ENAs show a quite broad enhancement of low energy emissions centered near the nose, but shifted somewhat toward the up-field longitude (to the starboard or right in this figure). However, there is a noticeable drop in intensity and number of pixels with S/N > 3, compared to the first five year averaged fluxes (McComas et al. 2014a). This is because of the global decrease in fluxes seen both by IBEX-Hi and –Lo over most of the IBEX epoch.

## 2.7 IBEX-Hi Maps of the Spectral Index

As we did in McComas et al. (2014a), we provide in Figure 15 the spectral indices calculated as linear fits to the measured fluxes in the five IBEX-Hi energy steps for each pixel in the sky and for each year from the Ram and Anti-ram maps. Also, as we did before, we forgo the full C-G correction and opt for the simpler (and model independent) process of just correcting the fluxes in each pixel for the effect of the spacecraft's ~30 km s$^{-1}$ speed, owing to the Earth's orbital motion. This process simply puts the observations in the Sun's inertial frame. Figure 16 provides similar maps including corrections for the time- and energy-dependent ENA survival probabilities.

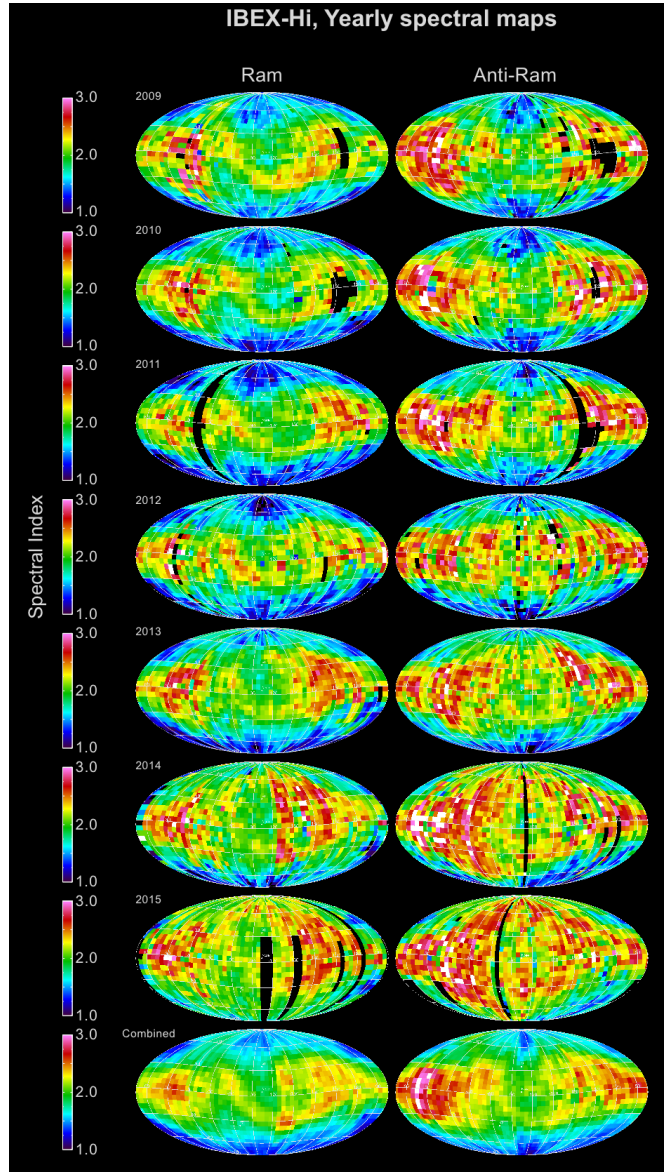

**Figure 15.** *Sky maps of energy spectral index from IBEX-Hi data (~0.5 to 6 keV) corrected to the solar frame.*

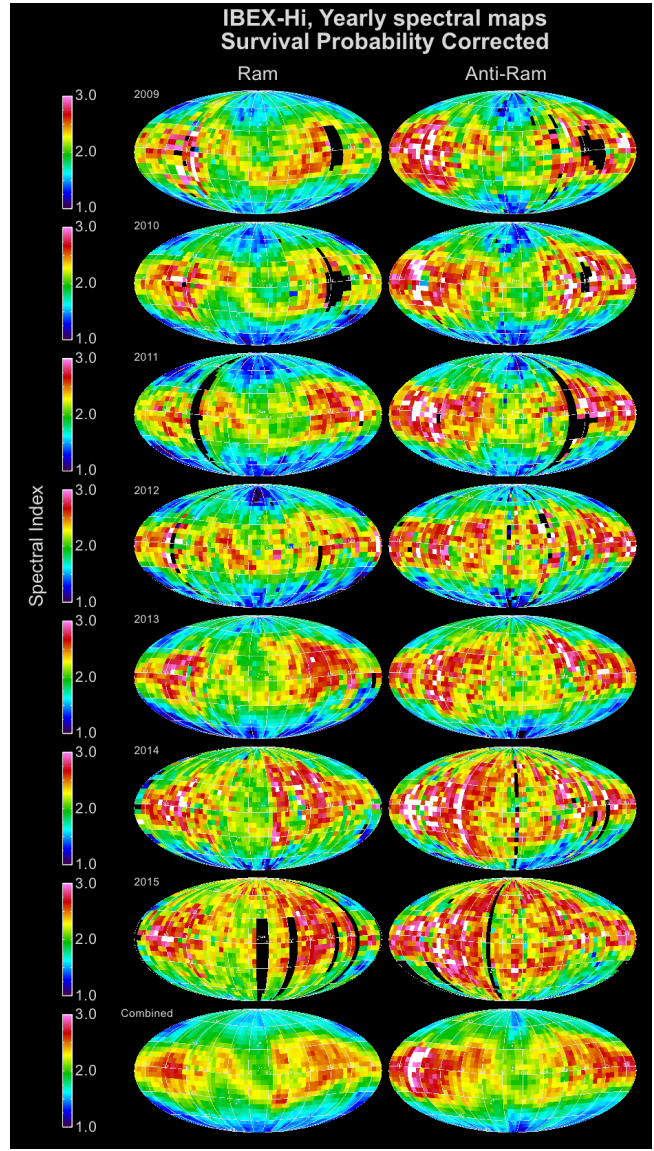

***Figure 16.*** *Same as for Figure 15, but survival probability corrected.*

Figures 15 and 16 show several important trends in the overall spectral indices over time. First, higher spectral indices that dominate the low- to mid-latitudes in the earlier years of IBEX data, extend to progressively higher latitudes over time. This is consistent with the breakdown of the large-scale circumpolar coronal holes that persisted through the prior solar minimum and the several year "recycle" time for the solar wind to populate the inner heliosheath and Ribbon and propagate back into 1 AU. It is also interesting that the spectral index has increased over time in nearly all directions in the sky.

The IBEX Ribbon is partially visible in these spectral maps, especially in the earlier years and near the center of the plots (nose), with a spectral index somewhat larger than the immediately surrounding regions. However, spectral indices across the maps, even in the higher latitude regions, have increased in the last two years to values of typically ~2-3. This reflects the disappearance of high energy PUIs associated with the fast SW. In the past, the IBEX energy spectra from high latitudes were generally less well fit by a single power law than at low to mid latitudes and tend to show an upward inflection around the middle of the IBEX-Hi energy (McComas et al. 2009c; Dayeh et al. 2012). The new data show larger indices and while the Ribbon can still be discerned near the nose, it is not as clearly visible as it was near the start of the mission. This may be yet another indication of distinctly different source location for the Ribbon compared to the GDF.

The spectral indices observed on the sides of these Mollweide projections show larger values in the two broad regions at low to mid latitudes that represent the port and starboard lobes of a large heliotail structure centered on opposite sides of the downwind direction (McComas et al. 2013b). In the last two years, the spectral index has increased above 2.5 at low latitudes as with most of the rest of the sky. However, the lower indices (~1.5-2) at higher latitudes on the tailward side seem to show the least temporal evolution in their spectral index. This may be due to the longer line-of-sight regions that contribute to these pixels and the (at least initially) fast solar wind flowing down through these northern and southern lobes. More detailed analyses of the temporal variations in the observed ENAs are taken up in Section 3, below.

## 2.8 Different Map Views Highlight Different Results

In this section, we highlight several alternate display formats for the integrated seven years of IBEX data, as we did previously for the five-year study. Figure 17 provides Mollweide projections oriented exactly opposite of those shown in Figure 12. That is, while the data is the same, the maps are centered on the opposite direction – the downwind instead of the upwind direction (McComas et al. 2015b). As first shown by McComas et al. (2013b), this perspective is ideal for examining the heliotail region, which is now roughly centered in the plots. In these plots, it is easy to see the region of enhanced flux coming from the downwind direction at low to mid energies, and especially the division of this enhancement, moving higher and lower latitudes (McComas et al. 2013b; Schwadron et al. 2014c; Zirnstein et al. 2016a) and the emergence of port and starboard lobes with very low emissions on the two sides at the two highest energies (McComas et al. 2013b).

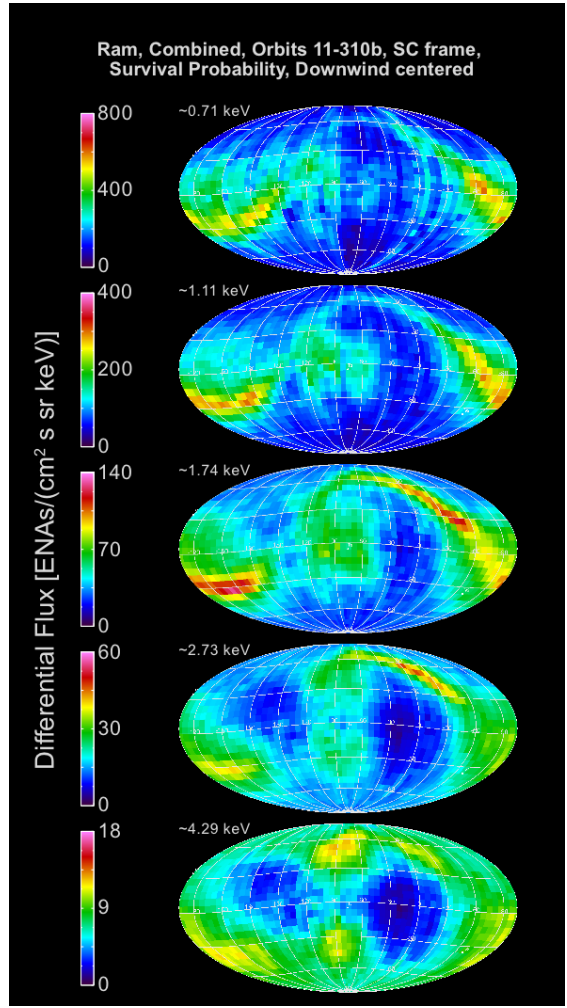

**Figure 17.** *Same data as in Figure 12, but centered on the exactly opposite (downwind) direction. This view is especially good for examining the heliotail, which is roughly centered in the plots (McComas et al. 2013b).*

Figures 18 and 19 provide additional perspectives, with Mollweide projections centered on the Ribbon (Funsten et al. 2013b) in the upwind (Figure 18) and downwind (Figure 19) hemispheres, which is approximately the direction of the external, interstellar magnetic field (McComas et al. 2009c; Schwadron et al. 2009). Recently, Zirnstein et al. (2016b) utilized the IBEX Ribbon center location as a function of ENA energy with MHD/kinetic modeling to precisely determine the pristine interstellar magnetic field properties, unperturbed by the heliosphere's presence, under the assumption of a secondary ENA source for the Ribbon. Zirnstein et al. (2016b) showed that each ENA flux map is uniquely constructed from ENAs originating from overlapping source regions in the OHS, as a function of ENA energy (see their Figure 1). Due to this unique coupling, the source of the IBEX Ribbon as a function of ENA energy outside the heliosphere is coupled to spatially varying regions of draped interstellar magnetic field. A comparative analysis of IBEX data and simulations were used to precisely determine the magnitude and direction of the pristine field far from the Sun. Since most Ribbon ENAs observed by IBEX originate in regions of space where the interstellar magnetic field is perturbed by the heliosphere, the draped field shifts the center of the Ribbon away from the pristine interstellar magnetic field direction along the B-V plane by ~8° towards the LISM inflow direction (Zirnstein et al. 2016b). Thus, Figures 18 and 19 are centered on the observed Ribbon center (219.2°, 39.9°), which is slightly offset from the pristine interstellar magnetic field direction ~(227°, 35°).

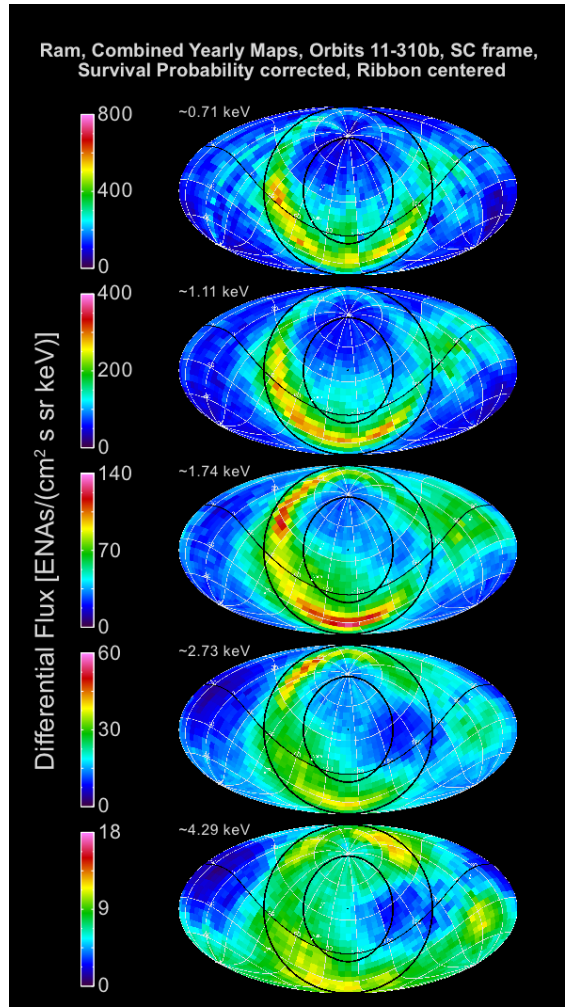

***Figure 18.*** *Mollweide projection of the seven-year combined IBEX-Hi ENA fluxes centered on the Ribbon toward ecliptic J2000 (219.2°, 39.9°) (Funsten et al. 2013b). The circles and curved line for the ecliptic plane (same in all panels) are there to guide the eye to differences at different energies.*

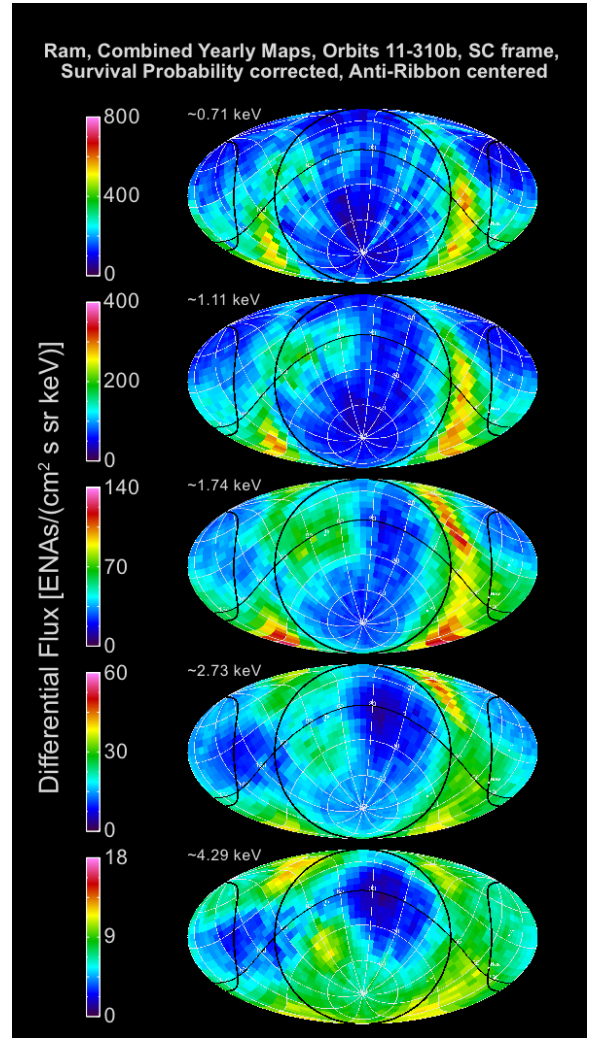

**Figure 19.** *Similar to Figure 18, but centered on the opposite direction – anti-parallel to the Ribbon center in the downwind hemisphere. As in Figure 17, the lines indicate the ecliptic plane and approximate boundary of the Ribbon. Because the Ribbon has a half cone angle of ~74.5° (Funsten et al. 2013b), it appears to emerge from the top and bottom of the plots in this projection, and connects through the vertical sections of the plots.*

In addition to showing data in ecliptic coordinates, we also provide the data in galactic coordinates. Figure 20 shows the ram, yearly-averaged, survival probability corrected data in galactic coordinates. In this projection, the maps are based on a Sun-centered observer, and the center of the plot is directed toward the galactic center. One can see that the Ribbon is significantly offset from the galactic plane, which runs horizontally across the center of the Mollweide projections in Figure 20.

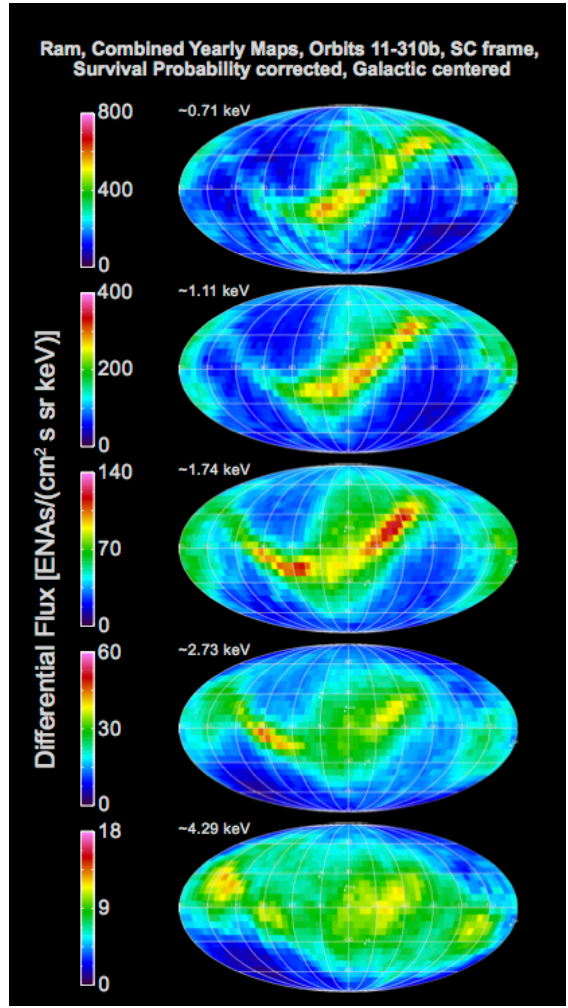

***Figure 20.*** *Mollweide projection of the seven-year combined IBEX-Hi ENA fluxes in galactic coordinates. We note that the raw IBEX data was re-binned into the pixels in galactic coordinates, so no interpolation was required.*

In Figure 21, we also provide the ram, yearly-averaged, survival probability corrected data in equatorial J2000 coordinates. In this projection, the north pole points parallel to the Earth's rotation axis, and the plot is centered on the vernal equinox. Here the Ribbon appears shifted in longitude toward the left side of the maps.

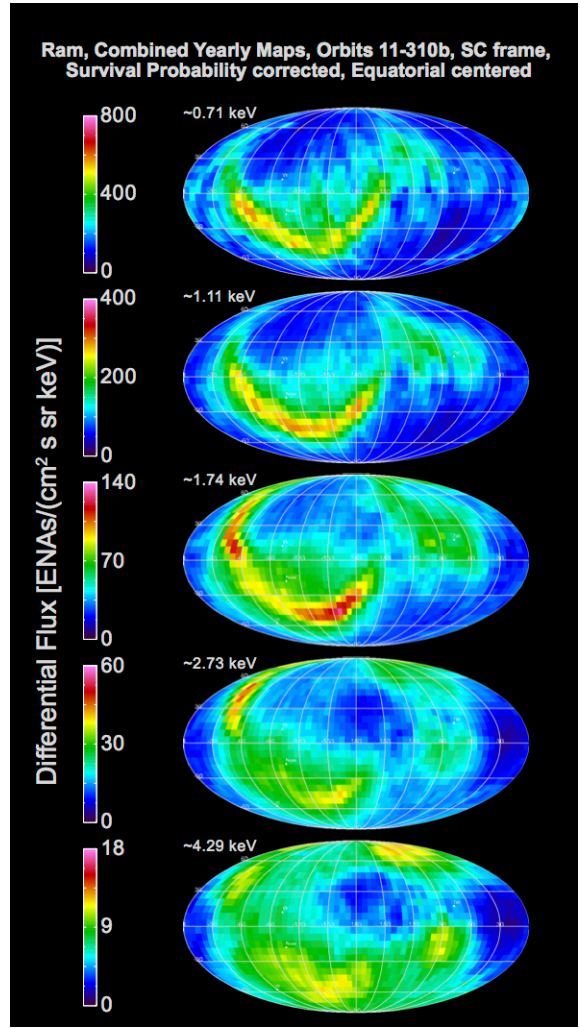

***Figure 21.*** *Mollweide projection of the seven-year combined IBEX-Hi ENA fluxes in equatorial J2000 coordinates. Similar to Figure 20, the raw IBEX data were re-binned into pixels in equatorial coordinates without interpolation.*

## 3. TIME VARIATIONS OVER SEVEN YEARS OF IBEX OBSERVATIONS

As the IBEX mission has progressed, we observed time variations in the ENA fluxes that reflect the temporal evolution of the global heliospheric interaction. The first two analyses of this type were McComas et al. (2010), which compared just the first two set of sky maps (2009A and 2009B), and McComas et al. (2012c), which examined time variations over the first three years (maps 2009A-2011B). This latter study introduced and compared the annual ram and anti-ram maps over three years; these maps do not suffer from the uncertainties introduced by the C-G correction, so they are better suited for the study of temporal variations.

McComas et al. (2012c) found that heliospheric ENA emissions had decreased from 2009 to 2011 with the Ribbon decreasing the most and at least part of the structure in the heliotail decreasing the least. These authors argued that the decreasing ENA flux was driven by a generally decreasing outward-traveling solar wind flux over the prior years and a several-year "recycle" time for solar wind ions to make it out into the inner heliosheath, become neutralized, and transit back to 1 AU where IBEX observes them. This argument also led McComas et al. (2012c) to predict the level of further reduction expected in the following two years on ENA measurements, based on the solar wind mass and momentum flux already observed at 1 AU at that time; these predictions were in fact confirmed by subsequent observations (McComas et al. 2014a).

Reisenfeld et al. (2012) used the fact that the IBEX observational geometry provides continuous viewing of the two polar regions to look for variations on time scales shorter than IBEX observations from other parts of the sky: six months for maps that require C-G corrections to compare or one year without invoking C-G corrections. These authors did not find significantly shorter time scale variations. However, they did confirm the steadily decreasing fluxes in both polar regions for the two-year period from December 2008 to February 2011. The same conclusion was reached by Dayeh et al. (2014), using the spectral indices from the polar ENA fluxes. Recently, Reisenfeld et al. (2016) analyzed the last several years of data, again from just the poles, and found an energy-dependent recovery in the fluxes, with the lowest energies recovering sooner, in contrast to the simple idea that the faster traveling ENAs should show a recovery sooner. These authors interpret this as the disappearance of fast solar wind at the poles during the recent solar maximum, causing the high energy ENA fluxes from these directions to continue decreasing.

McComas et al. (2014a) examined the first five years of IBEX data (maps 2009A-2013B), which was a long enough interval to begin to look for the effects of the ~11 year solar cycle and its time variable three dimensional structure (McComas et al. 1998; 2003; 2008). The five years of observations showed a general decrease in the ENA fluxes from 2009 to 2012 and a possible leveling off in 2013 over most of the sky. In the heliotail direction, however, no leveling off was observed and fluxes continued to fall in 2013. In addition, the Ribbon showed a more complex variation with a leveling off in the southern hemisphere and continued decline in the northern one. Overall, while not definitive, the results were consistent with a 2-4 year recycle time from most of the inner heliosheath. Longer times are required for the tail and the Ribbon, with a secondary ENA process in the outer heliosheath, suggesting that these regions are farther away. The Ribbon appears to be farther away in the north than in the south owing to the draping and compression of the external interstellar magnetic field.

For all analyses of time variations in our current study, we follow the same guidelines as in McComas et al. (2012c, 2014a), that is, we: 1) make comparisons only with non C-G corrected maps in order to avoid introducing additional uncertainties from such a correction, 2) compare ENA fluxes that have been corrected by their energy-dependent survival probabilities in transit to 1 AU, and 3) primarily compare annual sets of maps (ram and anti-ram) separately, so that the exact same viewing geometry exists for each pixel from year to year.

Figure 22 compares the seven annual ram maps for 2009-2015 of the ~1.1 keV ENAs to the timeline of solar wind dynamic pressure (white) and smoothed sun spot number (red). While the first five years of data show a continuous decrease, ENA fluxes in the last two years have flattened out (as we will see below, some have even started to recover). Due to the typical ~2-4 year recycle time for the closer parts of the heliosheath, this is consistent with the decline in SW dynamic pressure at 1 AU up to ~2010, and flattening in the following years. Over the second half of 2014 there was a steep and significant rise in the dynamic pressure, which we predict will soon be reflected in IBEX data as enhanced ENA emissions from the nearest regions of the inner heliosheath (see Section 4 - Discussion, below).

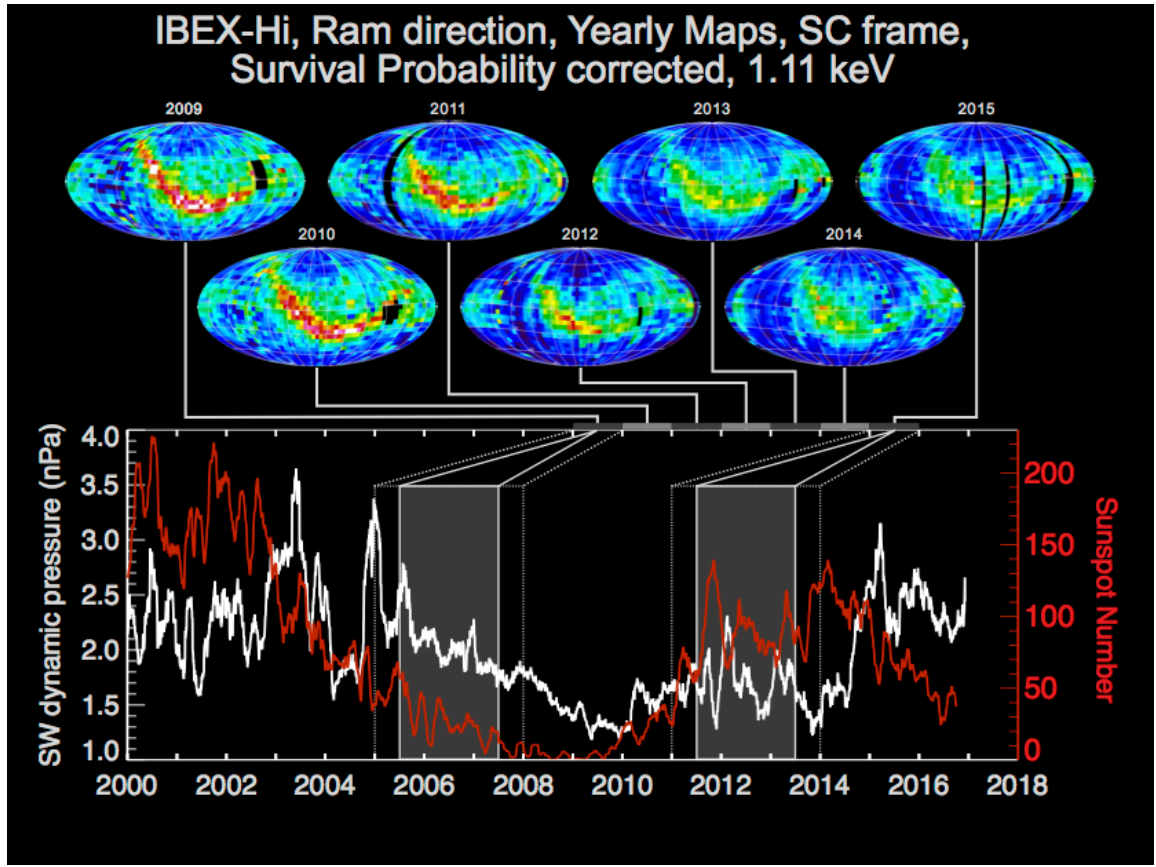

**Figure 22.** *IBEX ENA maps of survival probability corrected 1.1 keV ENAs (top) compared to the time series (bottom) of the solar wind dynamic pressure at 1 AU (white), and sunspot number (red). For typical "recycle" times across most of the sky of ~2-4 years (shaded for 2009 and 2016) and year-long maps (additional dotted lines), solar wind variations observed at any given time produce ENA emissions with this sort of multi-year time delay.*

In contrast to what was done in the three and five-year papers, in this study we provide more finely defined sub-regions of the heliospheric structure in order to better quantify time variations in the ENA fluxes. Figure 23 identifies and provides the temporal variations for eight separate regions, that characterize different parts of the outer heliospheric interaction: the core of the Ribbon, north and south poles, regions of relatively pristine GDF on the southern upwind and downwind sides, and tail portions from the port and starboard lobes and central down-tail region, including the northern and southern lobes. For this analysis, we used survival probability corrected ENA fluxes to remove the time-variable losses of ENAs on their transit in from the outer heliosphere to 1 AU. It is important to note that while the statistical error over such large regions are quite small (some error bars are hidden by the points), some additional systematic errors remain, owing to the imperfect background subtractions. The lowest energy (ESA2, yellow curves) fluxes are most susceptible to such effects and undoubtedly contain some additional background.

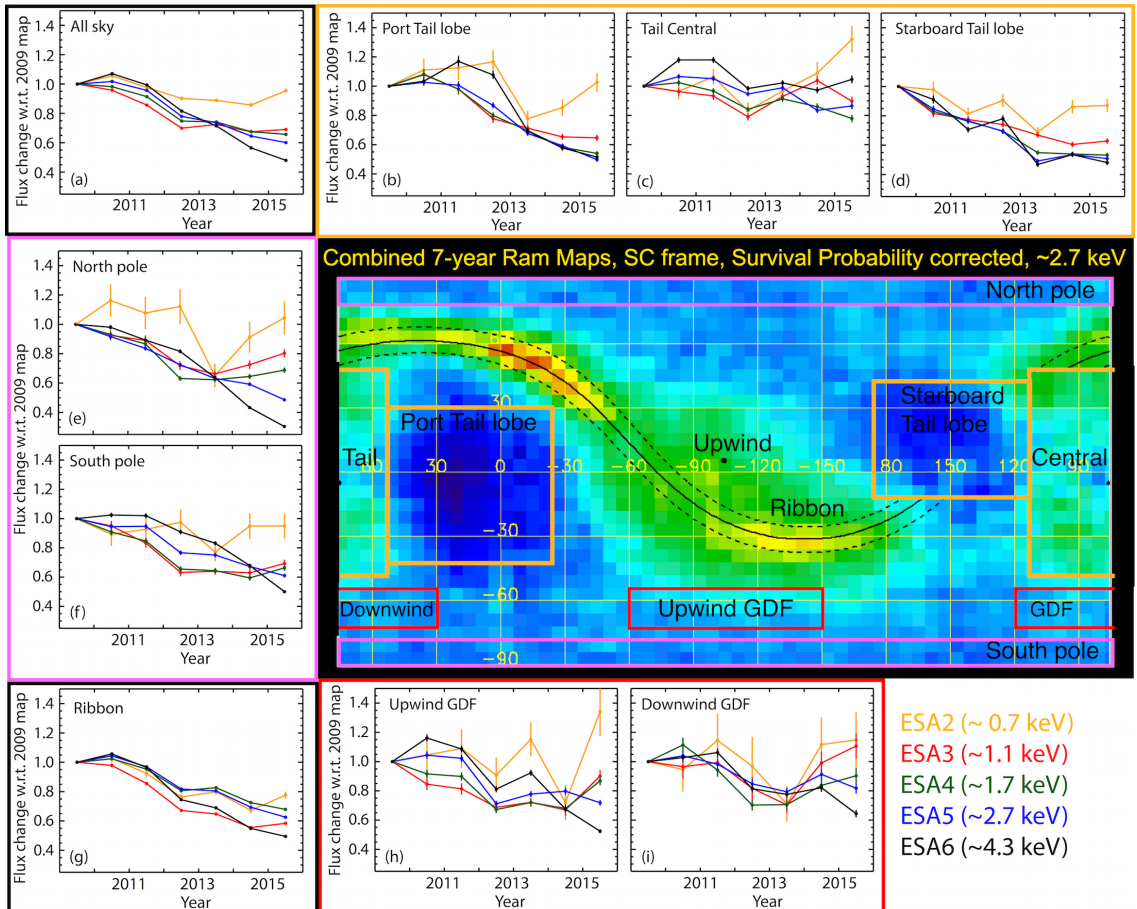

***Figure 23.*** *Combined 7-year ram map at 2.7 keV overlaid with lines identifying eight regions in the sky maps, surrounded by the quantitative time variations in each region (color coded and labeled). Energy passbands are also color-coded (bottom right) and*

*data is plotted along with statistical error bars, normalized to the 2009 fluxes observed in each region.*

The curves in Figure 23 tell an interesting and complex story, with different regions and energies displaying different temporal variations. At the top level, the all sky panel (top left) shows that ENA fluxes have generally dropped over the IBEX epoch, with lower energies flattening and even starting to recover around 2012-2013 and higher energies continuing to decrease. The polar regions are similar, consistent with that found by Reisenfeld et al. (2016), who interpreted the continued reduction at high energies as being due to the disappearance of fast solar wind at the poles during the recent solar maximum. The upwind and downwind GDF at relatively high latitudes show roughly flat emissions over the past few years with enhancements at the lower energies. These results are similar to those at the poles, but because they do not reflect just the fastest solar wind from the solar minimum polar coronal holes, they show a less pronounced relative reduction. Interestingly, the downwind GDF did not drop as much as the upwind GDF at the same southern heliolatitudes. This may be due to the thinner inner heliosheath on the upwind side compared to the downwind and therefore the downwind side is less sensitive to relatively short term variations in the solar wind flux.

The ENA fluxes from the port tail lobe have continued to generally decrease, but at a slowing rate, while the starboard lobe fluxes have leveled off. In our 5-year paper (McComas et al. 2014a), we predicted from the observed outgoing solar wind that "we would expect a leveling off in these fluxes from the slow solar wind heliotail lobes a couple years later than for the rest of the sky". Observations shown in Figure 23 confirm this prediction and are consistent with the greater integration lengths and "deeper" history of solar wind being sampled down the port and starboard tail lobes compared to the upwind direction. The differences in the recovery of fluxes between the port and starboard tail lobes show again that the heliosphere's interaction is highly asymmetric and not well described by simple symmetric models.

In contrast to the port and starboard lobes, the downwind tail direction and northern and southern lobes, collectively (central tail), have shown significantly less reduction over the entire IBEX epoch; this result likely indicates that ENAs from this unique direction span a longer range of distances and times back in the past and effectively average out more of the solar wind fluctuations over the past. Finally, the overall Ribbon fluxes show a general reduction over the IBEX epoch, with a flattening only in the last year or two. The detailed story of the time variations of the Ribbon fluxes is more complicated and taken up in detail below.

Figures 22 and 23 show that the variations in the ENAs coming in from both the Ribbon and the rest of the sky have changed over time, and these changes are a function of the ENA energy. To examine this more quantitatively, next we look at the time variations in the flux from the 1800 individual 6°x6° pixels in the sky. Figure 24 shows the flux variations for a typical 6° x 6° pixel in the ~1.7 keV energy channel. Taken across all seven years, a linear fit would indicate a relatively slow decrease (green fit line). However, in reality, the decrease was much more rapid and occurred only over 2009-2012 and since 2012 the fluxes have been relatively constant or increasing. Using the analysis shown in this figure for each pixel in the sky, we calculated all three time-variation slopes: a seven-year linear fit and separate linear fits for the years 2009-2012 and 2012-2015, where the latter fits are fixed at the 2012 value.

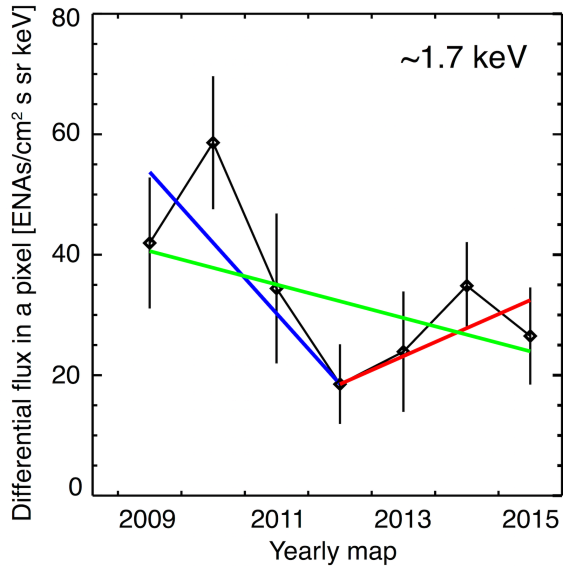

**Figure 24.** *Example of a 6°x6° pixel that shows the measured flux of ~1.7 keV ENAs for each year. This pixel is centered on ecliptic longitude and latitude (3°, 81°), near the north pole above the Ribbon knot. In contrast to the linear fit across all years (green), the time variations are better characterized separately in two epochs: from 2009-2012 (blue), which show more rapidly decreasing fluxes, and 2012-2015 (red), which show flat to slightly recovering fluxes in this and many other pixels. Both the latter fits are constrained to match at the measured 2012 value.*

Figure 25 compares sky maps of the three slopes in ENA flux for each pixel in the sky, as defined in Figure 24 (2009-2015, 2009-2012, and 2012-2015). The maps clearly show a single linear fit to the fluxes (left column) is no longer a good way to characterize the evolving ENA emissions. In contrast, most of the pixels show more rapid decreases (blue) from 2009 to 2012, while only some pixels at the lowest and highest energies increased. Then, from 2012 to 2015, most of the pixels at least partially recovered (red) at the lower energies. While a leveling off in ENA fluxes in 2013 was found in the prior McComas et al. (2014a) paper, and even predicted from a simple model two years prior to that (McComas et al. 2012c), the addition of the 2014-2015 data makes the situation clear. Not only did the fluxes generally level off at lower energies, but across much of the sky, lower energy ENA fluxes have increased since their low points around 2012.

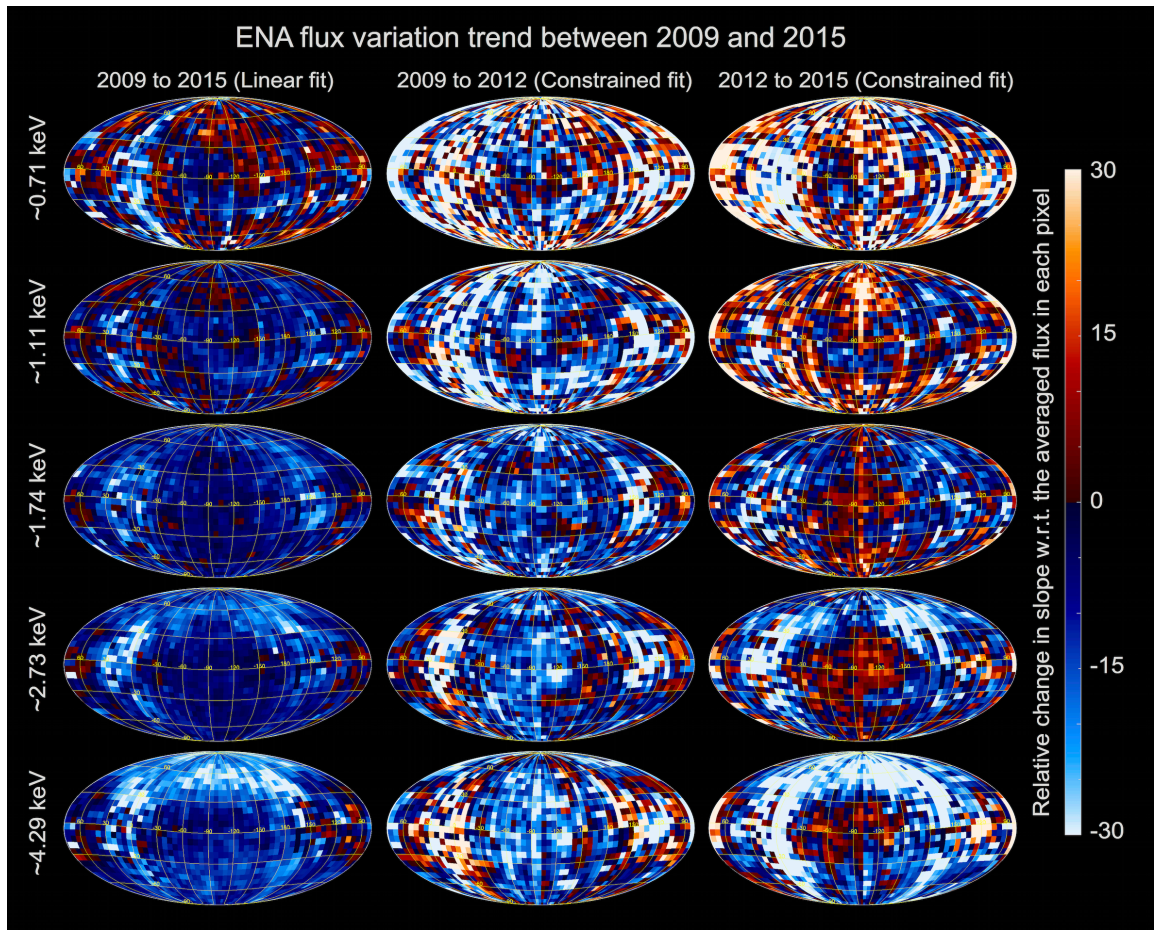

*Figure 25. ENA flux variation trend between 2009 and 2015. The left column shows results from a linear fit over all 7 years; in contrast, the middle and right columns show results from linear fits constrained to adjacent subintervals from 2009-2012 and 2012-2015, respectively.*

In contrast to the lower energies, the ENA fluxes have not increased since 2012 at higher energies of ~1.7 to ~4.3 keV, except in the upwind direction, which are closest to the Sun and most quickly recycle solar wind into inward propagating ENAs. Over the rest of the sky, and especially around the flanks, ENA fluxes continue to show strong decreases at higher energies. Interestingly, ENA fluxes from the central down-tail direction first appear to decrease from 2009 to 2012, but then increase from 2012 to 2015 at every ENA energy.

Another, complimentary type of analysis is shown in Figure 26. Here, we display observation time-weighted differences in absolute fluxes averaged over various ranges of years. The left column shows the difference between the average fluxes from the first three years and last two years of the data taken so far (2014-2015 vs 2009-2011). The second and third columns provide a similar analysis, but dividing the data approximately up into thirds (2012-2013 vs 2009-2011 and 2014-2015 vs 2012-2013).

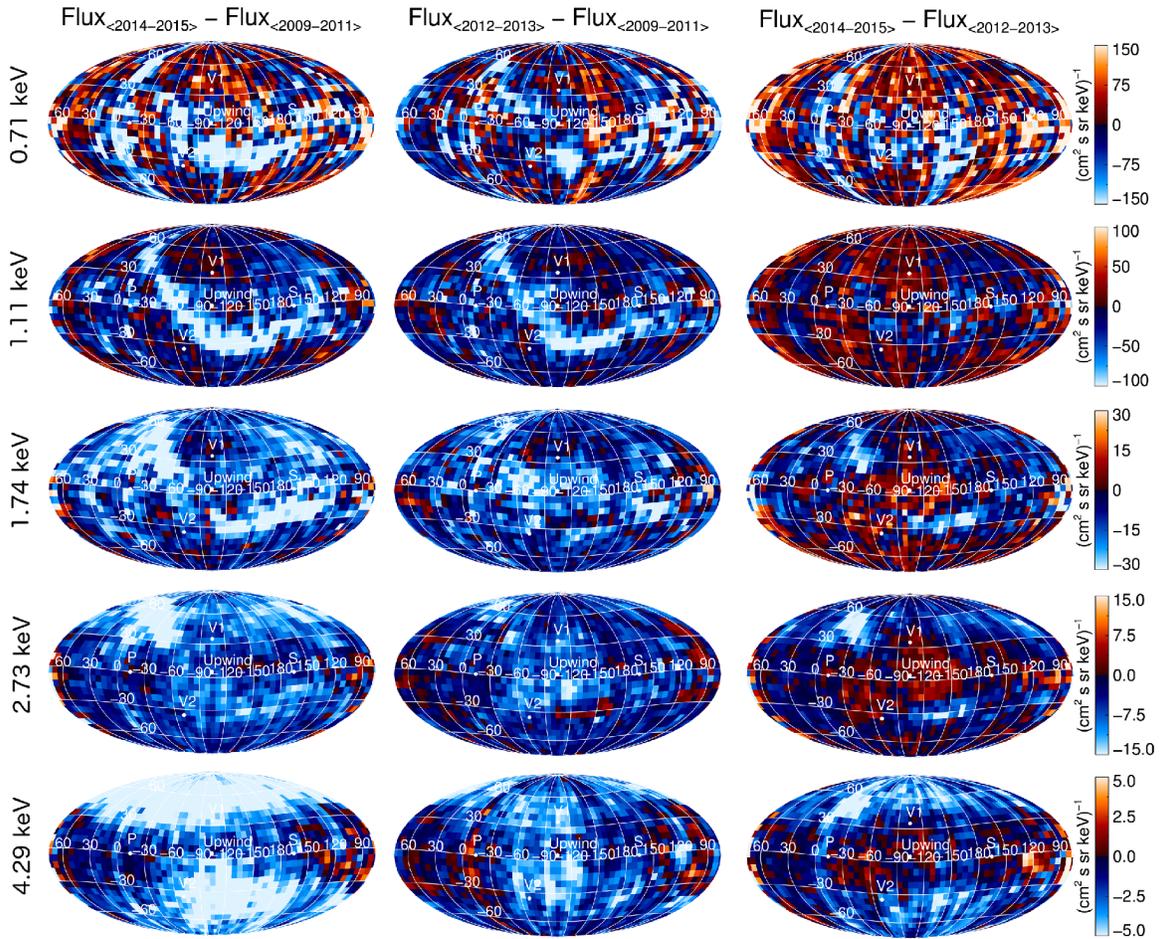

***Figure 26****. Differences between absolute ENA fluxes averaged over different sets of years: 2014-2015 minus 2009-2011 (left); 2012-2013 minus 2009-2011 (middle) and 2014-2015 minus 2012-2013 (right).*

Because Figure 26 shows changes in absolute instead of relative fluxes, it is a particularly good format for identifying changes in features that have different quantitative variations. The Ribbon stands out as decreasing far more than the rest of the sky at the lower energies at low- to mid-latitudes. While it is expected that the Ribbon shows the greatest absolute change in flux because it is the brightest feature in the sky, interestingly in the last two years of observations (2014-2015), the Ribbon appears to evolve quite differently than the surrounding GDF. The Ribbon at ~1.1 keV, for example, while decreasing with most of the rest of the sky from 2009 to 2013, continues to decrease in 2014-2015 as the surrounding fluxes begin to recover. This is also visible in the Ribbon at ~0.7 keV, although with more variations, and at ~1.7 keV in the southern portion of the Ribbon. This behavior is an indication of a Ribbon source that takes longer to process solar wind changes into ENA changes than in the surrounding GDF. At higher energies, the mid- and high-latitude fluxes decrease the most. Reductions of ENAs from the "knot" in the Ribbon (McComas et al. 2010) are most clearly seen in the upper left portions of the 1.7, 2.7, and 4.3 keV maps. Large reductions in the knot are consistent with the loss of fast polar coronal hole solar wind at mid-latitudes of ~50°-60° north. Comparison of the second and third columns indicate that the reduction in the knot was largest in 2014-2015. While the surrounding GDF at high latitudes also continues to decrease in 2014-2015, in the next few years we may see a recovery of the high latitude GDF, but a continuation of the drop in the Ribbon knot flux.

Overall, Figure 26 suggests a Ribbon source location that is quite different from the GDF ENAs arising from the rest of the sky. This provides additional evidence for a separate source region, not co-located with the ENAs arising from the inner heliosheath, and therefore supports some sort of secondary ENA source mechanism beyond the heliopause. This general type of source was initially proposed by McComas et al. (2009c) and subsequent studies have provided additional detailed secondary source mechanisms and quantified the expected ENA signals (Heerikhuisen et al. 2010; Chalov et al. 2010; Gamayunov et al. 2010; Schwadron & McComas 2013; Isenberg 2014; Giacalone & Jokipii 2015; Zirnstein et al. 2015a,b). Two major changes in the solar wind most strongly affect ENA intensities in Figure 26: 1) changes in solar wind dynamic pressure and 2) opening/closing of the polar coronal holes. The closing of the polar coronal hole is responsible for the drop in ENA intensity at mid- and high-latitudes at ~2.7 and 4.3 keV. The small recovery in solar wind output and dynamic pressure at 1 AU after 2010 is reflected in the inner heliosheath ENA fluxes ~2-4 years later (Figure 26, right column), but will likely not affect the Ribbon fluxes until a few years later.

As in our prior 5-year study (McComas et al. 2014a), here we separately examine fluxes from various sub-regions of the Ribbon. We found this examination to be important in our prior study because the Ribbon fluxes varied jointly in latitude and energy following the last solar minimum in a way that was consistent with the latitude dependent solar wind source around solar minimum (McComas et al. 2012c). That correlated variation showed the strongest Ribbon emissions at low energies from the solar wind at low latitudes, at high energies from the fast solar wind at high latitudes, and at a range of intermediate energies from mixtures of fast and slow winds at intermediate latitudes.  McComas et al. (2012c) argued that the latitudinal ordering with strong ENA emissions for lower energies at low latitudes (slow solar wind) and for higher energies at higher latitudes (fast solar wind) demonstrated a quite direct recycling of the solar wind ions into Ribbon ENAs, such as that provided by a secondary ENA source process. In this study, we retain the sub-regions identified in the 5-year study (McComas et al. 2014a), as shown in Figure 27, and extend the prior analysis by adding ENA fluxes for 2014 and 2015 for each of the various sub-regions in Figure 27.

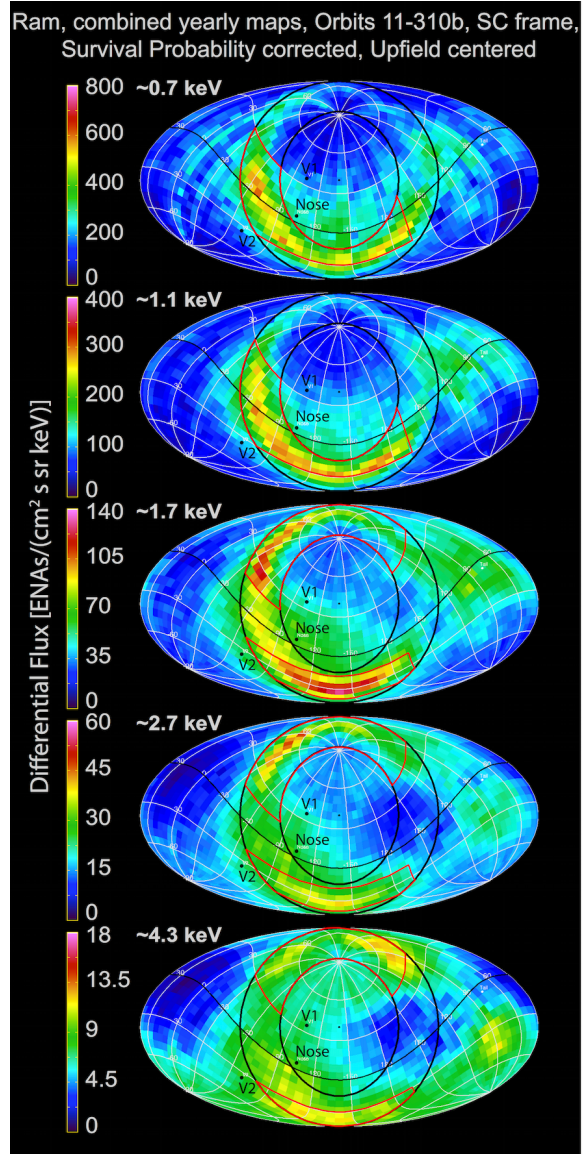

**Figure 27.** *Regions of the Ribbon identified in the prior the 5-year study (McComas et al. 2014a), overlaid on the full 7-year set of observations.*

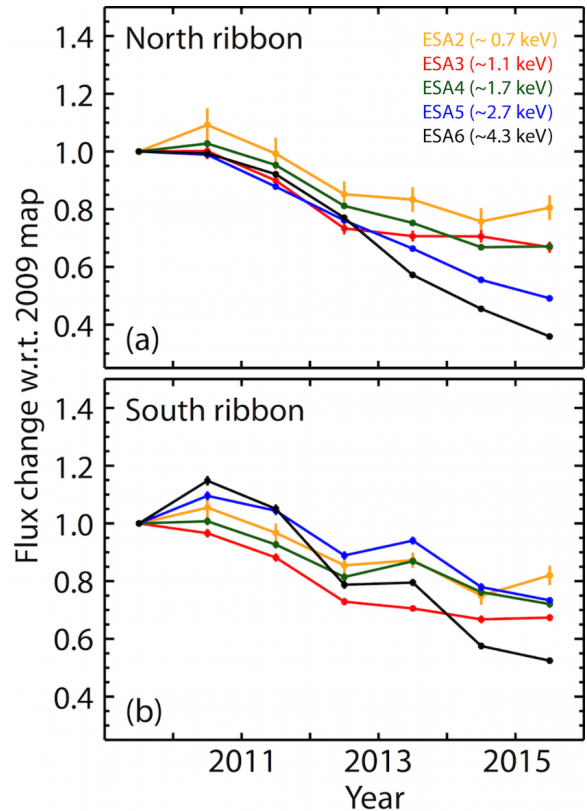

**Figure 28.** *ENA fluxes from the latitude-dependent Ribbon regions specified in McComas et al. (2014a) and shown in Figure BB.*

Figure 28 extends the time evolution of ENAs in the various sub-regions of the Ribbon identified in the prior 5-year study. While fluxes have generally leveled off in the lowest three energy bands since ~2012, the fluxes at the highest energy (~4.3 keV) have continued to drop, as have the fluxes at the next highest energy (~2.7 keV) in the north, and to a lesser extent in the south. It is interesting to note that after the apparent flattening in the southern Ribbon in 2012-2013, ENA fluxes again dropped. In contrast, in the northern Ribbon, the flattening has been later and less dramatic. Such differences again point to significantly different distances to the Ribbon source in the north and south, both because 1) the Ribbon extends to higher latitudes in the north and 2) the southern portions of the Ribbon are largely in the upwind direction, and thus compressed by the inflowing interstellar plasma and relatively strong external field, which preferentially compresses the upwind side of the heliosphere in the south (McComas et al. 2009c; Schwadron et al. 2009; Opher et al. 2009; Pogorelov et al. 2011; McComas & Schwadron 2014). Moreover, the northern and southern Ribbon sub-regions are influenced by solar wind at different latitudes. The northern sub-region reflects Ribbon ENAs influenced by solar wind output at higher latitudes than the south, as well as reflecting the differences between the northern and southern polar coronal hole output (Karna et al. 2014; Sokół et al. 2015; Reisenfeld et al. 2016). Note the similar behavior of the northern polar flux in Figure 23 to that in the Ribbon, except that the northern Ribbon fluxes largely continue to decrease or at most level off.

Another way to look at the evolution of the Ribbon fluxes over time is shown in Figure 29. The top row shows the original three years (2009-2011), which McComas et al. (2012c) used to discover the strong correlation between latitude and energy for the peak Ribbon fluxes: low energies coming for slow solar wind at low latitudes, high energies coming from fast solar wind at high energies, and a broader range of intermediate energies coming from intermediate latitudes, consistent with the solar wind structure around solar minimum. These authors went on to predict that, eventually, this latitudinal ordering of the Ribbon fluxes should break down as the solar wind speed is no longer well-ordered at solar maximum and this lack of order would work its way through the heliosphere. Further, they suggested that the time it takes for these changes to be reflected in the Ribbon would strongly constrain the possible source location and mechanism.

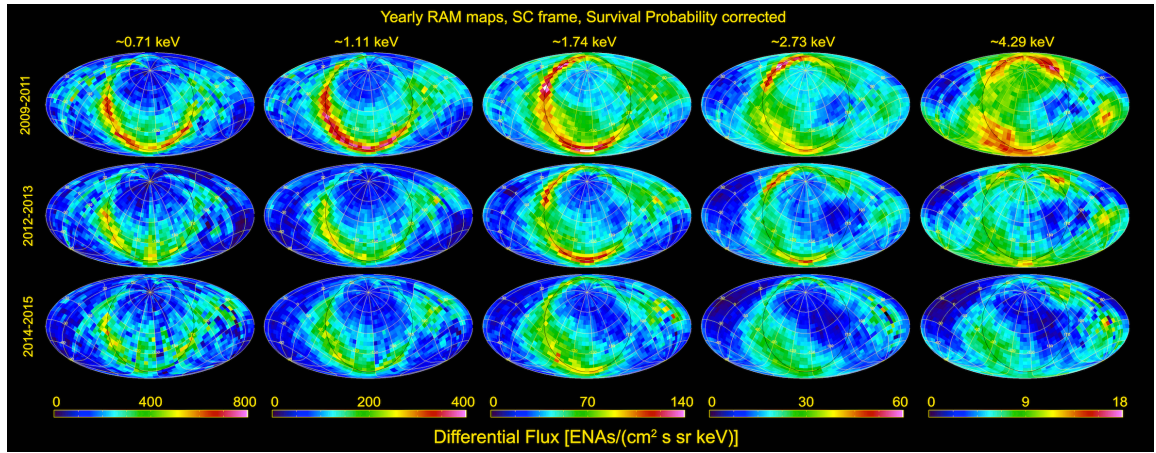

***Figure 29.*** *Survival probability corrected, yearly ram maps. Maps are time-averaged from 2009-2011 (top), 2012-2013 (middle), and 2014-2015 (bottom) for all ENA energies. Note that at the beginning of the mission, the Ribbon flux portrayed latitudinal and energy-dependent ordering related to the fast-slow solar wind structure (McComas et al. 2012c). In the last few years, however, this ordering has broken down, reflecting solar maximum conditions.*

The second and third rows of Figure 29 (2011-2012, middle row and 2014-2015, bottom row) clearly shows the progression away from latitudinal ordering of the peak ENA emissions, as predicted by McComas et al. (2012c). In 2014-2015, the latitudinal ordering of the high energy ENAs appears to be nearly broken. Let us analyze the "knot" at 2.7 keV, which is at latitude ~60° in the northern hemisphere. Figure 26 shows it to be decreasing significantly from ~2013 to 2015, and Figure 29 shows the latitudinal ordering of the Ribbon to be nearly absent in 2014-2015. Considering that observations of the northern polar coronal hole fractional area (Karna et al. 2014) and IPS solar wind observations (Sokół et al. 2015) show the fast solar wind to disappear at latitudes <60° in ~2011, this yields an estimate for the recycle time of ~3-4 years. This is slightly shorter than the predicted recycle time expected for 2.7 keV secondary ENAs from outside the heliopause (~4-6 years; Zirnstein et al. 2015b). The shorter time compared to simulation suggests this may not reflect the final solar maximum conditions yet, the ENA source region is closer (note the unexpected early crossing of the heliopause by Voyager 1), or the charge-exchange lifetime of pickup ions outside the heliopause is shorter.

## 4. DISCUSSION

In this study, we have examined the global ENA observations from the full IBEX mission, both adding the sixth and seventh years of observations (2014-2015) for the first time and providing several small improvements to the earlier five years of data. This study provides the documentation needed for researchers to be able to readily use the IBEX data provided in the associated data release. With the addition of the sixth and seventh years of data, we now have over a half solar cycle of observations; these provide substantially more information about the variations of fluxes of ENAs from various regions of the outer heliospheric interaction and allow us to begin to dissect the sources and time histories of these regions separately.

The single largest discovery of the IBEX mission was that of the completely unanticipated and unpredicted IBEX Ribbon, and not surprisingly, the single most important question raised by this discovery is that of its origin. Over seven years, numerous, quite different ideas have been advanced, studied, and debated. Over time, many researchers have come to see a source beyond the heliopause via some sort of secondary ENA process as the most likely explanation. In this study, with the benefit of two additional years of observations we now argue that a secondary ENA source mechanism should be adopted as the primary and most likely explanation of the Ribbon.

Figure 30 summarizes the change in the Ribbon between ENAs observed over 2009-2011 (left in pairs of maps, black) and 2014-2015 (right in pairs of maps, red), and compares them to schematic diagrams of the solar wind latitudinal structure (lower left). During solar minimum (2007-2010), there was fast solar wind from large circumpolar coronal holes in both the north and the south, slow solar wind at low latitudes, and a mixture of fast and slow winds at mid-latitudes, as is typically seen around solar minimum conditions (McComas et al. 1998; 2008). This led to peak ENA fluxes in the IBEX Ribbon ~3 keV at high latitudes (the Ribbon only extends to high latitudes in the north), ~1 keV at low latitudes, and a broader range of intermediate energies at intermediate latitudes. Subsequently, this simple latitudinal ordering broke down in the approach to solar maximum. While the ENA maps from 2012-2013 didn't show much change, those from 2014-2015 (right column in this figure) are markedly different and now largely reflect a source that is not latitudinally ordered and which has a variety of ENAs, indicating a variety of speeds in its source solar wind, at all latitudes.

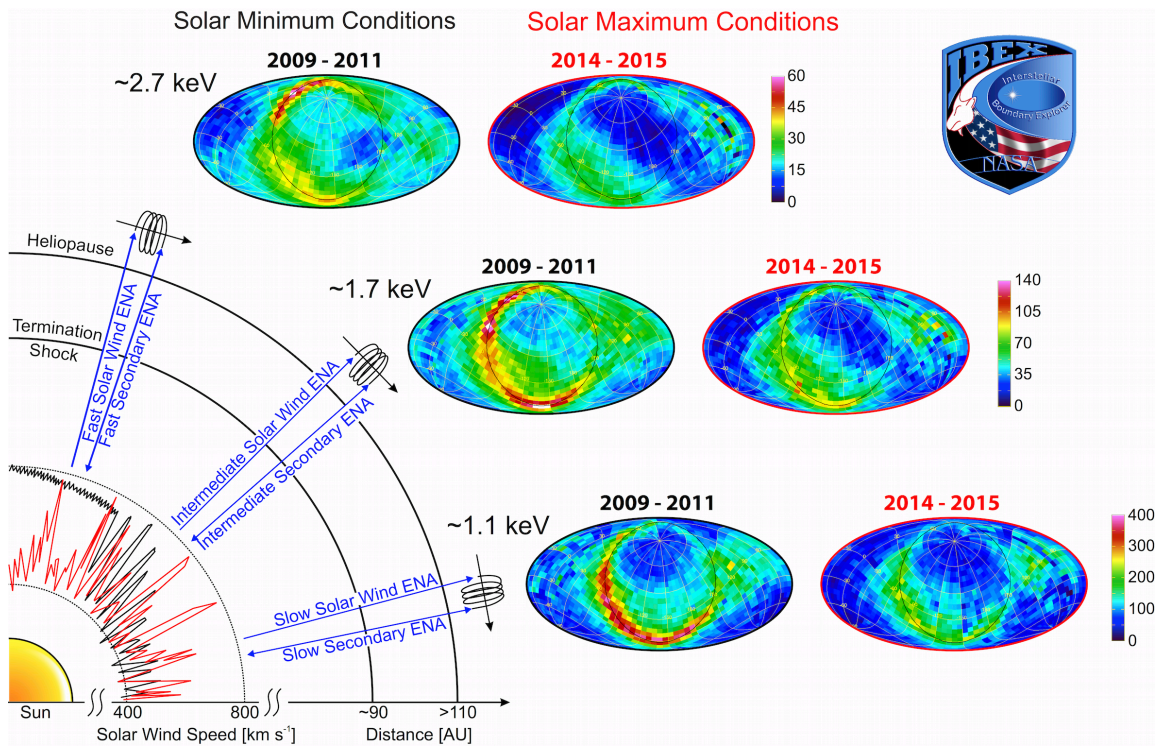

***Figure 30.*** *Combined IBEX ENA data and schematic diagram highlighting the differences between Ribbon emission reflective of solar minimum (left set of Mollweide projection maps, black) and those indicative of the breakdown of solar wind-latitude order in the approach to solar maximum (right set of maps, red). Together, these demonstrate the response of the Ribbon to solar minimum (fast solar wind at high latitudes, slow wind at low latitudes) and solar maximum (slow to intermediate solar wind speeds at all latitudes) conditions, and the recycling time between them.*

The predicted (McComas et al. 2012c, 2014a), and now confirmed change from the latitudinal ordering from solar minimum to the disordering of solar maximum of the Ribbon confirms that a very direct recycling process for outflowing solar wind ions to create the Ribbon must be occurring. The most direct way to recycle solar wind into heliospheric ENAs with the same latitudinal ordering (or lack of ordering) is via the secondary ENA process (McComas et al. 2009c; Heerikhuisen et al. 2010; Chalov et al. 2010; Gamayunov et al. 2010; Schwadron & McComas 2013; Isenberg 2014; Giacalone & Jokipii 2015; Zirnstein et al. 2015a,b). Other recent lines of evidence also support the conclusion that a secondary ENA source of the Ribbon is most likely. For example, work by Swaczyna et al. (2016a) analyzed the parallax of Ribbon ENAs and found a radial distance to the Ribbon source of ~140+84/-38 AU, which is consistent with an origin in the region close to, but beyond the heliopause – exactly where the secondary source is expected to peak (see, e.g., Zirnstein et al. 2015a, 2016b). The energy-dependent position of the Ribbon (Funsten et al. 2013b) was also found to be directly linked to the latitudinal ordering of the solar wind under the secondary ENA hypothesis (Swaczyna et al. 2016b).

In the present study, we have shown that the latest IBEX observations indicate a disparity in the temporal evolution of the Ribbon and surrounding GDF. Over the first five years of observations (McComas et al. 2014a), there was evidence that the GDF was generally dimming over time, while parts of the Ribbon showed evidence for a leveling or slight increase. However, in this study, with the benefit of 7 years of IBEX observations, we are now able to differentiate between the Ribbon and other ENA time variations. The data show that the Ribbon at ~1 keV continues to decrease in intensity over 2014-2015, while the GDF from most directions surrounding the Ribbon was increasing. This behavior is consistent with the recovery in solar dynamic pressure now being reflected in the GDF, but a longer recycling time for the solar wind that produces the Ribbon ENAs. Such a difference clearly adds strong support to the explanation of a secondary ENA source for the Ribbon beyond the heliopause.

The significant differences in the evolution of the Ribbon compared to the GDF not only reflects their different sources, but also indicates that most of the fluxes contributing to what we call the GDF come from a similar source, very likely in the inner heliosheath. Earlier studies pointed out the potentially significant contribution of secondary ENAs from outside the heliopause to the ~keV GDF signal (i.e., secondary ENAs coming from directions away from the Ribbon; Izmodenov et al. 2009; Opher et al. 2013; Desai et al. 2014; Zirnstein et al. 2014). While this may be true at lower ENA energies (Desai et al. 2014; Zirnstein et al. 2014), the fact that there is an observable difference in the evolution of the Ribbon and surrounding GDF of at least 1 year suggests that most of the GDF signal originates from the inner heliosheath. We also note that the GDF observed by IBEX are most likely primary ENAs. All primary ENAs, produced either in the supersonic solar wind or the inner heliosheath, have mean free paths >10,000 AU in the inner heliosheath, and thus will likely not ionize until passing through the denser outer heliosheath/LISM, where the mean free path is significantly smaller (~100-1000 AU). Therefore, the results of this study conclude that the GDF is mostly comprised of a primary ENA source from the inner heliosheath (at ~keV energies), and the Ribbon from a secondary ENA source outside the heliopause.

Given the conclusions of this study, we now have a basis for predicting future ENA fluxes that should be observed over IBEX's continuing extended missions and ultimately by the follow-on Interstellar Mapping and Acceleration Probe (IMAP) – see the Heliophysics Decadal Survey (National Research Council 2013; see also McComas et al. 2011c; Schwadron et al. 2016b). These predictions reflect different time frames, from the nearest term based on already existing solar wind observations at 1 AU, to very long term variations in the solar dynamo and solar wind output.

The solar wind observed at 1 AU over the past several years is already being processed through the heliosphere. With typical speeds of ~400 km s$^{-1}$, the solar wind takes about one year to reach the termination shock at ~100 AU and similar energy ENAs (~1 keV) take about another year to come back from these distances. Processing times in the inner heliosheath can span from a year or two near the nose to much longer times back toward the tail due to the longer line-of-sight and different plasma flows (e.g., Zirnstein et al. 2016y). Secondary ENA mechanisms beyond the heliopause have typical re-ionization timescales of roughly two years (depending on the ENA energy), and primary/secondary ENA travel times of one or two years between 1 AU and the Ribbon source. Finally, the solar wind data in the ecliptic plane can be used as a proxy for the full three-dimensional solar wind as McComas et al. (2008, 2013a) showed that the mass and momentum fluxes vary globally, so ecliptic values are representative of these parameters at all solar latitudes on average.

As shown in Figure 22, over the second half of 2014 there was a rapid and significant rise in dynamic pressure, from values of ~1.5 nPa for the couple years before to ~2.5 nPa for 2015-2016. Such a large, sustained increase is both unusual and fortuitous as it provides an excellent opportunity to examine the propagation of such a large solar wind change (providing a unique signal) through the outer heliospheric processes that ultimately recycle the solar wind and embedded pickup ions back into heliospheric ENAs. At least toward the nose and direction of maximum heliosheath pressure ~20º southward (McComas & Schwadron 2014), we predict that this dramatic increase in solar wind dynamic pressure will soon be reflected in IBEX data as enhanced ENA emissions from these regions – most likely in the 2017-2018 time-frame – with changes from the Ribbon and flanks/near tail following by a couple years.

Beyond the next few year timeframe, the solar cycle effects should again be observable in the IBEX data. In particular, the current cycle (24) has completed solar maximum and the polar coronal holes should be rebuilding toward more solar minimum-like conditions. Under these conditions, fast solar wind is re-emerging at high latitudes and smaller scale coronal holes are coalescing and forming the next large circumpolar coronal holes and latitude ordered slow and fast solar wind that characterize about half of the solar cycle around solar minimum. As this happens, fast wind and latitudinal ordering will again fill the northern and southern higher latitude heliosphere and ultimately return as latitudinally-ordered ENAs as observed over the first 5 years of the IBEX mission. We predict that this ordering will first emerge in the high latitude GDF ENAs on the upwind side of the heliosphere and probably in the south before the north owing to its closer distance. After that, the ordering should progress further at lower latitudes, back away from the nose and flanks, and ultimately be seen again in the Ribbon fluxes, which we expect will lag the GDF by a couple of years.

Finally, the future longer term fluxes will reflect not just the most recent 1 AU data and ~11 year solar cycle variations, but also even longer term trends in the solar wind output, driven by long term variations in the solar dynamo. The deep and prolonged solar minimum between solar cycles 23 and 24 and the activity in solar cycle 24 have differed significantly from previous cycles during the last 100 years (Schwadron et al. 2011b, 2014b; McComas et al. 2013x). The fast solar wind was somewhat slower, while the solar wind in general was less dense, cooler, and had significantly lower momentum and mass fluxes (McComas et al. 2008). In addition, the solar wind had significantly weaker heliospheric magnetic fields (Smith & Balogh 2008) compared to earlier cycles within the space age. As the activity level of solar cycle 24 rose, the mass flux of solar wind remained extremely low (McComas et al. 2013a) and the magnetic flux of the solar wind remained at much lower levels than observed at previous solar maxima in the space age (Smith et al. 2013). As a result, solar cycle 24 is the weakest solar maximum of the space age, which continues the anomalous trends observed in the deep cycle 23-24 minimum. Conditions during the cycle 23-24 solar minimum were similar to conditions at the beginning of the Dalton Minimum (Goelzer et al. 2013). These recent changes suggest that the next solar minimum may continue to show a decline in sunspot numbers, and cause further reductions in magnetic flux and solar wind particle flux.

Alternately, the significant increase in dynamic pressure observed at 1 AU in the second half of 2014 could be the end of the weaker solar wind and a resumption of solar wind conditions more representative of those observed through the earlier portions of the space age. In either case, the heliosphere will process these long-term variations in the solar wind and pickup ions that become embedded in it.

IBEX continues to be a remarkable mission of exploration and discovery. With seven full years of observations we now see the solar cycle variations in the ENAs processed both in the inner heliosheath and beyond, in the secondary ENA Ribbon source in the very local interstellar medium, beyond the heliopause. The next several years continue to promise new insights and opportunities to continue to mature our understanding of the outer heliosphere, how it is driven by the solar wind from inside, and how it interacts with the local interstellar medium beyond. Ultimately, even more exciting discoveries beckon when even higher sensitivity and resolution observations are available from the planned IMAP mission.

*Acknowledgements*. We gratefully thank all of the outstanding IBEX team members who have made this mission such a wonderful success. This work was carried out as a part of the IBEX project, with support from NASA's Explorer Program and Polish National Science Center grant 2015/18/M/ST9/00036. We also acknowledge solar wind ram pressure and sunspot number data collected together in the OMNI database (ftp://spdf.gsfc.nasa.gov/pub/data/omni). Work at Los Alamos was performed under the auspices of the US Department of Energy.

**Appendix A.** Mapping of specific source files at the ISOC to figures shown in this study.

| Figure | Description | Folders |
|---|---|---|
| 1 | A maps (first half years, aka "odd" maps), SC frame | hvset_map1<br>hvset_map3<br>hvset_map5<br>hvset_map7<br>hvset_map9<br>hvset_map11<br>hvset_map13 |
| 2 | B maps (second half years, aka "even" maps), SC frame | hvset_map2<br>hvset_map4<br>hvset_map6<br>hvset_map8<br>hvset_map10<br>hvset_map12<br>hvset_map14 |
| 3 | A (odd) maps, C-G corrected | hvset_cg_map1<br>hvset_cg_map3<br>hvset_cg_map5<br>hvset_cg_map7<br>hvset_cg_map9<br>hvset_cg_map11<br>hvset_cg_map13 |
| 4 | B (even) maps, C-G corrected | hvset_cg_map2<br>hvset_cg_map4<br>hvset_cg_map6<br>hvset_cg_map8<br>hvset_cg_map10<br>hvset_cg_map12<br>hvset_cg_map14 |
| 5 | Combined maps, C-G corrected | hvset_cg_single |
| 7 | A (odd) maps, C-G and Survival Probability corrected | hvset_cg_tabular_map1<br>hvset_cg_tabular_map3<br>hvset_cg_tabular_map5<br>hvset_cg_tabular_map7<br>hvset_cg_tabular_map9<br>hvset_cg_tabular_map11<br>hvset_cg_tabular_map13 |
| 8 | B (even) maps, C-G and Survival Probability corrected | hvset_cg_tabular_map2<br>hvset_cg_tabular_map4<br>hvset_cg_tabular_map6<br>hvset_cg_tabular_map8<br>hvset_cg_tabular_map10 |

| | | hvset_cg_tabular_map12<br>hvset_cg_tabular_map14 |
|---|---|---|
| **9** | Combined maps, C-G and Survival Probability corrected | hvset_cg_tabular_single |
| **10** | Ram, Yearly, SC frame, Survival Probability corrected | hvset_tabular_ram_year1<br>hvset_tabular_ram_year2<br>hvset_tabular_ram_year3<br>hvset_tabular_ram_year4<br>hvset_tabular_ram_year5<br>hvset_tabular_ram_year6<br>hvset_tabular_ram_year7 |
| **11** | Anti-ram, Yearly, SC frame, Survival Probability corrected | hvset_tabular_antiram_year1<br>hvset_tabular_antiram_year2<br>hvset_tabular_antiram_year3<br>hvset_tabular_antiram_year4<br>hvset_tabular_antiram_year5<br>hvset_tabular_antiram_year6<br>hvset_tabular_antiram_year7 |
| **12** | Ram, Combined years, SC frame, Survival Probability corrected | hvset_tabular_ram_single |
| **13** | Anti-ram, Combined years, SC frame, Survival Probability corrected | hvset_tabular_antiram_single |
| **14** | Combined years, SC frame, Survival Probability corrected | lvset_h_tabular_single |
| **15** | Combined years, inertial frame | hvset_cg_ram_single<br>hvset_cg_antiram_single |
| **16** | Combined years, inertial frame, Survival Probability | hvset_cg_tabular_ram_single<br>hvset_cg_tabular_antiram_single |
| **20** | Ram, Combined years, Inertial frame, Survival Probability corrected, Galactic centered | hvset_tabular_ram_galactic_single |
| **21** | Ram, Combined years, Inertial frame, Survival Probability corrected, Equatorial centered | hvset_tabular_ram_equatorial_single |

**Note.** Figures 17-19, 27, and 28 utilize data from Figure 12. Figures 22-24 utilize data from Figures 10 and 12. Figures 25, 26, 29, and 30 combine fluxes over different time periods. The combining equations can be found online at the IBEX Data Release 7 website:

http://ibex.swri.edu/ibexpublicdata/Data_Release_7/description.html

## Appendix B. Updated Survival Probability Corrections for IBEX-Hi and IBEX-Lo

For this study, we updated survival probabilities for the majority of the IBEX mission. Updates were made for orbits 8 through 286b due to changes in the composite Lyman-alpha time series from LASP (Laboratory for Atmospheric and Space Physics at the University of Colorado - http://lasp.colorado.edu/lisird), which are used to calculate the radiation pressure and photoionization rates. The changes are small for the first few years of IBEX observations and increase slightly over time compared to the previous values. Survival probabilities for orbits 256a to 311b were also updated to include a new set of solar wind speed data from the interplanetary scintillation (IPS) observations for 2015.

After the solar activity maximum in 2012-2013, when the slow and dense solar wind flows spread nearly from pole to pole, the solar wind started to reorganize again toward the standard bi-modal structure typical for solar activity minima (e.g., McComas et al. 1998). We took this change of the solar wind structure as a function of latitude into account in the calculation of the survival probabilities of the H ENAs inside the heliosphere. The survival probabilities of H ENAs against the interactions with solar wind and solar EUV radiation were calculated following the methodology presented in McComas et al. (2012c, 2014a).

The most effective ionization process for the ENAs observed by IBEX is charge exchange with solar wind protons (see Figure 31), so we require the solar wind speed and density out of the ecliptic plane. These were reconstructed following the model developed by Sokół et al. (2013) from the *in situ* in-ecliptic solar wind measurements compiled in the OMNI data base (King & Papitashvili 2005) and observations of IPS conducted by the Institute for Space-Earth Environmental Research (ISEE) at Nagoya University in Japan (Tokumaru et al. 2012). The survival probabilities reached a minimum around orbit 180, during the maximum of solar activity in 2012, when the ionization rates were the highest; thereafter they started to increase in concert with the decrease of solar activity. The variations in the northern and southern hemispheres are slightly different, which is due to the differences in the solar wind structure between the two hemispheres.

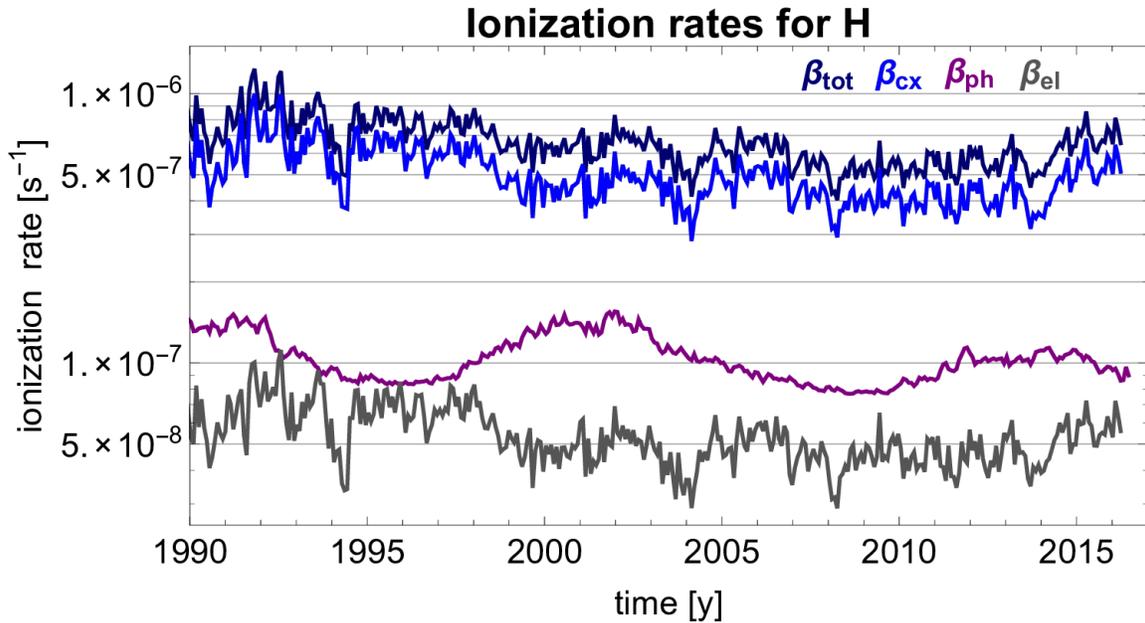

*Figure 31. Ionization rates for H in the ecliptic plane. The total ionization due to the three largest ionization processes is illustrated by the dark blue line ($\beta_{tot}$); separately, they are ionization from charge exchange ($\beta_{cx}$, blue), photoionization ($\beta_{ph}$, purple), and ionization due to impact with solar wind electrons ($\beta_{el}$, gray).*

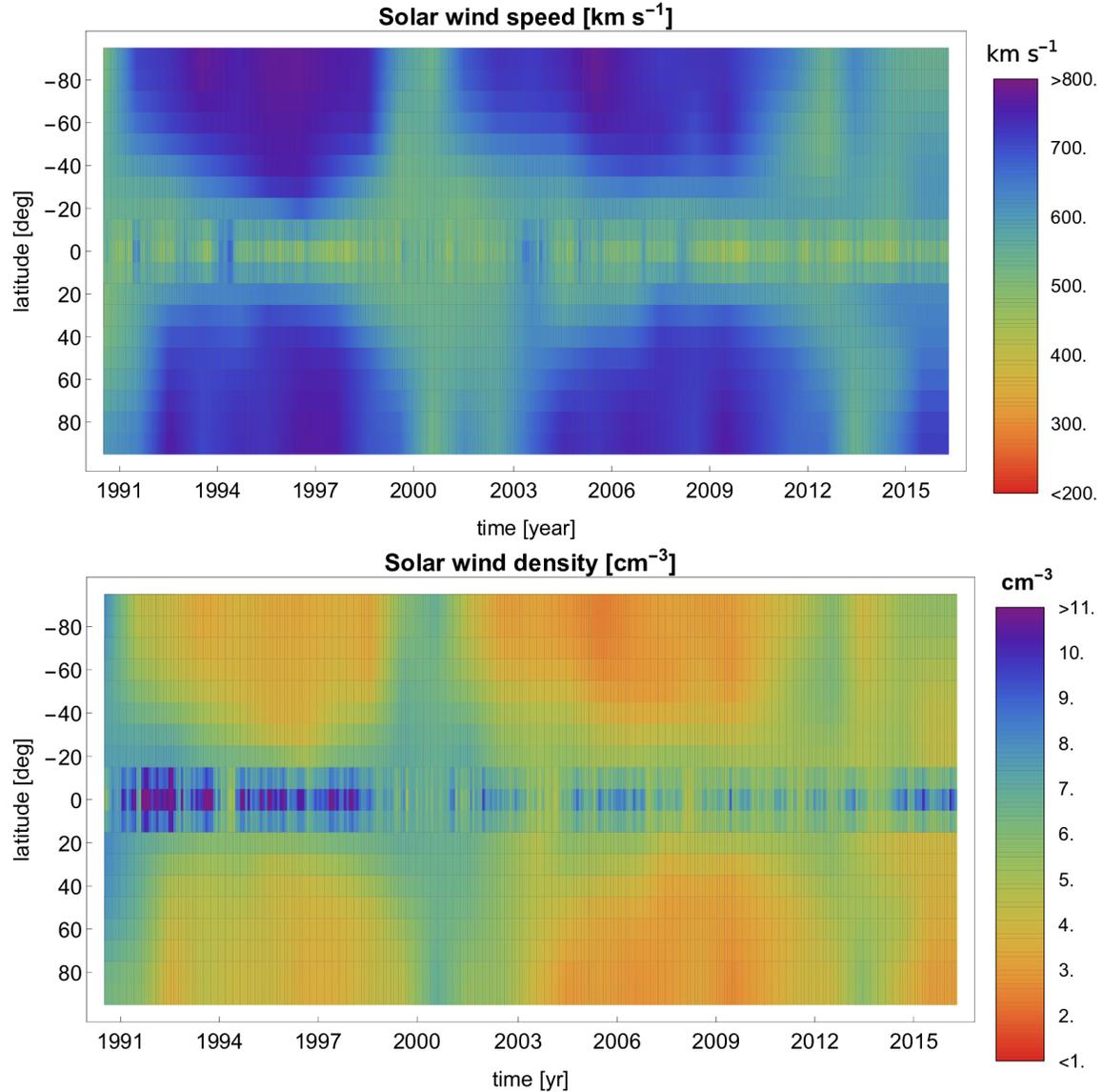

*Figure 32. Maps of solar wind speed and density as a function of time and heliolatitude reconstructed following the model described in Sokół et al. (2013).*

To reconstruct the global distribution of the solar wind speed (Figure 32, top panel) from the IPS observations using the computer assisted tomography method (CAT; Asai et al 1998; Jackson et al 1998; Kojima et al 1998), information of solar wind density fluctuations (ΔNe) is needed. One of the two CAT analyses assumes an empirical relation between solar wind speed and density fluctuations, ΔNe, while the other only uses speed estimates derived from multi-station IPS observations. Other versions do not assume such a model, but use two data sets; g-value data derived from single-station measurements and speed estimates from multi-station measurements. The g-value represents integration of ΔNe along the line-of-sight, and the IPS speed estimate is a convolution integral of the actual speed and ΔNe along the line-of-sight (Tokumaru et al. 2011, 2012). In the calculation of the survival probabilities we used the solar wind speed derived from the CAT analysis, which used both g-value and speed data (see more in Sokół et al 2013, 2015).

The solar wind density (Figure 32, bottom panel) is calculated from solar wind invariants in heliolatitude using the solar wind thermal advection energy flux (Le Chat et al. 2012), as presented in Appendix B in McComas et al. (2014a) and discussed by Sokół et al. (2015). The invariant is calculated from the in-ecliptic measurements, which together with the solar wind speed as a function of latitude, enable us to calculate the latitudinal solar wind density structure (see Equation B2 in McComas et al. 2014a).

In the survival probability calculation for H atoms we include the radiation pressure that competes with solar gravity. As in our previous study, we do this using a model from Tarnopolski & Bzowski (2009), with the total Lyman-alpha flux intensity obtained from the composite Lyman-alpha series provided by LASP.

Photoionization is of secondary importance for H atoms (Figure 31). Here we calculate it as before, integrating over the solar EUV irradiance measured by TIMED (Woods et al. 2005) and a hierarchy of solar EUV proxies, following Bzowski et al. (2013), and including the most recent data.

The time-variable survival probabilities for ~1 keV H ENAs observed in the ecliptic plane and towards the north and south poles are illustrated in Figure 33. The steps/jumps in the time series for the poles from orbit 232a to orbit 237a and subsequently from orbit 270a to orbit 279b in Figure 33 (see also Figure 6) are due to the changes in the spacecraft spin axis pointing to +/- 5° above and below the ecliptic plane, executed to facilitate interstellar neutral gas observations (e.g., Leonard et al. 2015; Bzowski et al. 2015; Möbius et al. 2015a; McComas et al. 2015b). The survival probabilities after orbit 294b were calculated using the latitudinal solar wind speed and density structure in latitude frozen in time due to the lack of more recent data; the most recent information about the solar wind structure out of the ecliptic plane is available up to the middle of 2015 (see details of the model construction in Sokół et al. 2013).

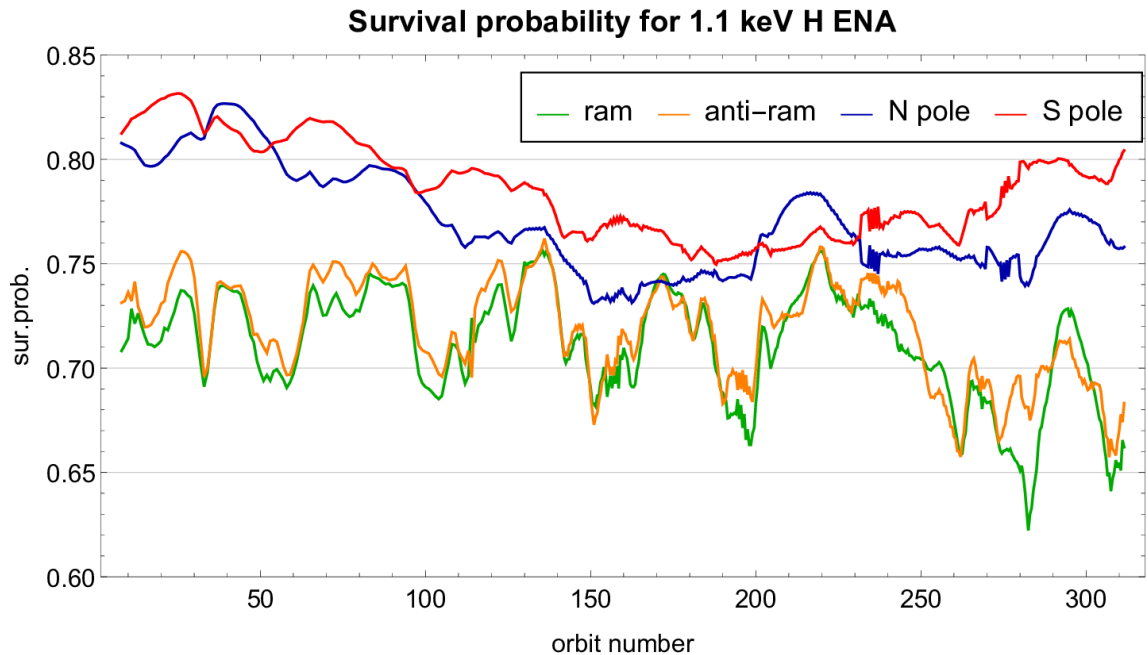

*Figure 33. Survival probabilities for H ENAs for the 1.1keV energy passband. Shows in-ecliptic pixel in the ram and anti-ram direction, and polar pixels towards north and south.*